\documentclass[10pt,twocolumn,letterpaper]{IEEEtran}
\usepackage{flushend}
\usepackage{graphicx}
\usepackage{epsfig}
\usepackage{latexsym}
\usepackage{amsfonts}
\usepackage{here}
\usepackage{rawfonts}
\usepackage[latin1]{inputenc}
\usepackage[T1]{fontenc}
\usepackage{calc}
\usepackage{url}
\usepackage{enumerate}
\usepackage{color}
\usepackage[tbtags]{amsmath}
\usepackage{amssymb}
\usepackage{upref}
\usepackage{times}
\usepackage{amsmath}
\usepackage{cite}

\newcommand{\ms}{\;\;}

\newcommand{\mycaption}[1]{\vspace*{-1.6em}\caption{#1}\vspace*{-1.0em}}

\newtheorem{lemma}{{\bf Lemma}}

\newcommand{\qed}{\nobreak \ifvmode \relax \else
  \ifdim\lastskip<1.5em \hskip-\lastskip
  \hskip1.5em plus0em minus0.5em \fi \nobreak
  \vrule height0.75em width0.5em depth0.25em\fi}

\newcounter{step}
\newlength{\totlinewidth}
  {\end{list}%
  \rule{\linewidth}{1pt}}
\newcounter{substep}

  {\end{list}}

\newlength{\aligntop}
\newlength{\alignbot}
\makeatletter
\renewenvironment{align}{%
  \vspace{\aligntop}
  \start@align\@ne\st@rredfalse\m@ne
}{%
  \math@cr \black@\totwidth@
  \egroup
  \ifingather@
    \restorealignstate@
    \egroup
    \nonumber
    \ifnum0=`{\fi\iffalse}\fi
  \else
    $$%
  \fi
  \ignorespacesafterend%
  \vspace{\alignbot}\par\noindent
}
\makeatother

\IEEEoverridecommandlockouts

\author{Manav R. Bhatnagar,~\IEEEmembership{Member,~IEEE} 
\vspace*{-2.9em}%
  \thanks{\textbf{Corresponding author:} Manav R. Bhatnagar is with
    Department of Electrical Engineering, Indian Institute of Technology Delhi, Hauz Khas, IN-110016 New Delhi, India,
    email: \protect\url{manav@ee.iitd.ac.in}.}%
}\date{}

\title{Decode-and-Forward Based Differential Modulation for Cooperative Communication System with Unitary and Non-Unitary Constellations\vspace*{-0.25em}}

\begin{document}

\maketitle
\vspace*{-1em}
\begin{abstract}
In this paper, we derive a maximum likelihood (ML) decoder of the differential data in a decode-and-forward (DF) based cooperative communication system utilizing \emph{uncoded} transmissions. This decoder is applicable to complex-valued unitary and non-unitary constellations suitable for differential modulation. The ML decoder helps in improving the diversity of the DF based differential cooperative system using an erroneous relaying node. We also derive a piecewise linear (PL) decoder of the differential data transmitted in the DF based cooperative system. The proposed PL decoder significantly reduces the decoding complexity as compared to the proposed ML decoder without any significant degradation in the receiver performance. 
Existing ML and PL decoders of the differentially modulated uncoded data in the DF based cooperative communication system are only applicable to binary modulated signals like binary phase shift keying (BPSK) and binary frequency shift keying (BFSK), whereas, the proposed decoders are applicable to complex-valued unitary and non-unitary constellations suitable for differential modulation under uncoded transmissions. We derive a closed-form expression of the uncoded average symbol error rate (SER) of the proposed PL decoder with $M$-PSK constellation in a cooperative communication system with a single relay and one source-destination pair. An approximate average SER by ignoring higher order noise terms is also derived for this set-up. It is analytically shown on the basis of the derived approximate SER that the proposed PL decoder provides full diversity of second order. In addition, we also derive approximate SER of the differential DF system with multiple relays at asymptotically high signal-to-noise ratio of the source-relay links. It is shown by simulations that the proposed PL decoder in the differential DF cooperative system with more than one relay also achieves the maximum possible diversity.    
\end{abstract}
\vspace*{0.0em} 
\section{Introduction}
Multiple-input multiple-output (MIMO) technology, proposed approximately a decade ago, revolutionized research in the field of wireless communications. By installing multiple antennas at the transmitter, benefits like diversity gain and spatial multiplexing can be achieved as compared to the single-input single-output (SISO) system~\cite{alamo98,tarok00}. It is proposed in literature~\cite{sendo03a,nosra04} that a relaying node can cooperate with a source node in order to support the transmissions of the source node to the destination. Therefore, the cooperative system utilizes distributed antennas to realize a virtual MIMO system which can provide benefits of a collocated MIMO system~\cite{lanem04}. 

There are two main protocols, namely amplify-and-forward (AF) and decode-and-forward (DF), proposed for the cooperative communication systems~\cite{lanem03}. In the AF protocol, the relaying node scales the received data before transmitting it to the destination node in order to satisfy a power constraint over the total transmit power. The destination requires knowledge of the channel coefficients of all links involved in cooperation for decoding the data of the source. Whereas, in the DF protocol, the relay first decodes the data transmitted by the source and then retransmits the decoded data to the destination. The destination requires knowledge of the channel gains of the source-destination and relay-destination links for decoding the data of the source. Since the relay \emph{cannot} decode the data perfectly, therefore, erroneous relaying causes significant error floor in the performance of the destination receiver. This is the primary reason why the AF based cooperation has been explored in much more detail as compared to the DF based system~\cite{jing06,jing07,song10}. 

Differential modulation is useful for the cooperative system as it enables the destination to decode the data of the source without any channel knowledge of the links involved in the cooperation. Therefore, the differential modulation is helpful in improving the spectral efficiency of the cooperative communication system~\cite{zhao05a,zhao05b,zhao07,himso05,chen06,fang09}.
Differential modulation for a symbol-wise DF based cooperative system utilizing uncoded transmissions with one source-destination pair, a single relay, and binary phase shift keying (BPSK) constellation is proposed for Rayleigh fading channels in~\cite{zhao05a,zhao07} and Nakagami-$m$ channels in~\cite{zhao06,zhao08}. 
A maximum-likelihood (ML) decoder for differentially modulated \emph{binary} frequency shift keying (BFSK) signal transmitted through multiple orthogonal regenerative relays using the symbol-wise DF protocol in the uncoded cooperative communication system is found in~\cite{chen06}. This ML decoder considers the possibility of erroneous transmission from the relay terminal and maximizes the probability density function (p.d.f.) of the received data in the destination terminal. In this way, it improves the diversity order of the DF based differential cooperative system~\cite{chen06}. A low complexity sub-optimal piecewise linear (PL) decoder of the \emph{binary} data is also obtained in~\cite{chen06}, which performs close to the ML decoder. Unfortunately, the ML and PL decoders of the differential DF cooperative system obtained in~\cite{chen06} are not applicable to the higher order complex-valued unitary and non-unitary constellations which are required for increasing the data rate of the wireless communication system. 
In literature~\cite{cui09}, differential modulation is also used in bidirectional relaying using $M$-PSK constellation, however, no ML or \emph{low complexity} PL decoders are obtained for this set-up. 

In this paper, we consider a cooperative system with one source-destination pair and a single \emph{unidirectional} relay for simplicity. Our main contributions are as follows. 1) We derive an ML decoder for the DF based differential cooperative system utilizing \emph{complex-valued unitary and non-unitary constellations}, and uncoded transmissions. 2) In order to significantly reduce the computational complexity in decoding of the differential data, we derive a PL decoder for complex-valued unitary and non-unitary constellations. 3) We derive a closed-form expression of approximate uncoded symbol error rate (SER) of the proposed PL decoder with $M$-PSK constellation. 4) It is analytically proved that the proposed PL decoder in the differential DF cooperative system with one source-destination pair and a single relay achieves second order diversity. 5)~An expression of the approximate uncoded SER of the differential DF cooperative system with \emph{multiple} relays under asymptotic condition of error-free source-relay links is derived and it is shown by simulation that the proposed PL decoder achieves the maximum possible diversity. 

The rest of this paper is organized as follows: In Section~II, the system model is introduced. Section~III derives optimal and low-complexity sub-optimal decoders of the differentially modulated unitary and non-unitary constellations in the DF based uncoded cooperative system with a single relay. In Section~IV, analytical performance analysis of the DF based differential cooperative system with a single relay and $M$-PSK constellation is performed. Differential modulation for the DF cooperative system with multiple relays is studied in Section~\ref{sec:mult}. The simulation and analytical results are discussed in Section~VI. Section~VII concludes the article. The article contains one appendix.
\section{System Model}
\begin{figure}[t!]
  \begin{center}\hspace*{0.0em}\vspace*{0.0em}
    \psfig{figure=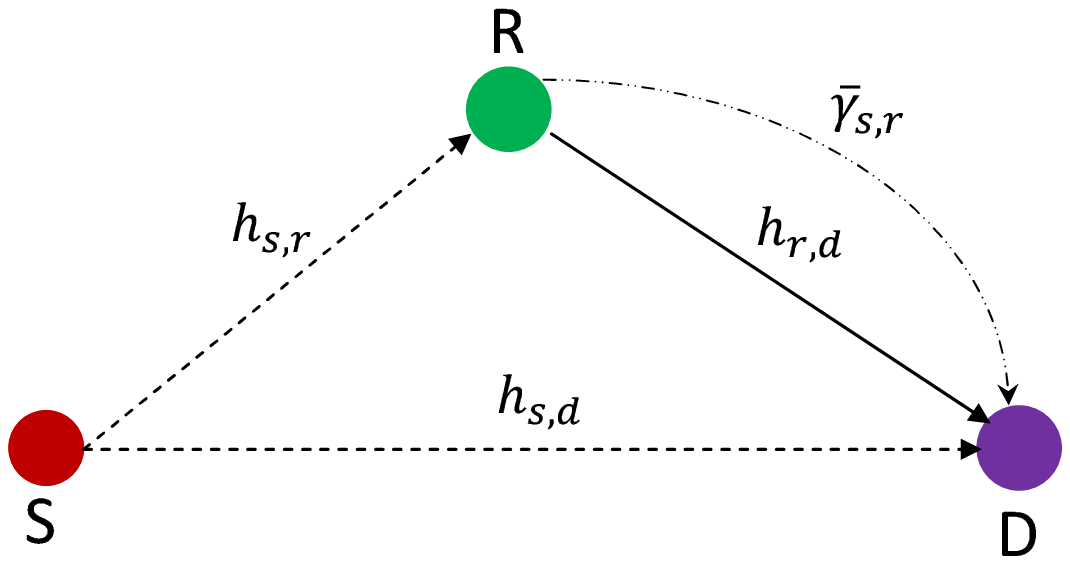,height=1.5in,width=2.5in}
    \vspace*{0.5em}
    \mycaption{Cooperative system with a single relay.}
    \label{fig:coopbd}
    \vspace*{-0.0em}
  \end{center}
\end{figure}

Let us consider a cooperative system containing one source, one destination, and a single relay as shown in Fig.\!~\ref{fig:coopbd}. Each node contains one antenna and it can either transmit or receive the data at a time. 
The transmission from the source to the destination is performed in two orthogonal phases. In the first phase, the source broadcasts \emph{uncoded} data to the relay and the destination. The relay demodulates the data of the source in symbol-wise manner and transmits the demodulated symbols to the destination in next phase. 
The source remains silent in the second phase in order to maintain orthogonality between the transmissions~\cite{lanem03}. 
The destination decodes the data of the source by utilizing an ML decoder in the second phase. We assume a differential cooperative system, where the relay and the destination do not require knowledge of the channel coefficients of the source-relay, relay-destination, and source-destination links for decoding of the data transmitted by the source. 

Let in the $n$-th time interval, the source needs to transmit a symbol $x[n]\in\mathcal{A}$, where $\mathcal{A}$ is a \emph{unit-norm} $M$-PSK constellation containing the following points: $\left\{x_1,x_2,x_3,...,x_M\right\}$.
Before transmission of $x[n]$, the source performs the following differential encoding: 
\begin{align}
\label{eq:diffencode}
v[n]=v[n-1]x[n],\ms n=1,2,3,...,
\end{align}
where $v[0]=1$ is an initialization symbol. As $\left|x[n]\right|^2=1$, therefore, $\left|v[n]\right|^2=1$. The data received in the destination during the first phase in the $n$-th time interval will be
\begin{align}
\label{eq:destinationdata}
y_{s,d}[n]=h_{s,d}v[n]+e_{s,d}[n],
\end{align}
where $h_{s,d}$ represents the circular complex Gaussian channel gain of the source-destination link with zero mean and $\sigma^2_{s,d}$ variance, and $e_{s,d}[n]$ is the signal-independent complex-valued additive white Gaussian noise (AWGN) with zero mean and $N_{s,d}$ variance.
During the first phase, the data received in the relay in the $n$-th time interval can be written as
\begin{align}
\label{eq:r1r2recdata}
y_{s,r}[n]=h_{s,r}v[n]+e_{s,r}[n],
\end{align}
where $h_{s,r}$ denotes the circular complex Gaussian channel gain of the source-relay link with zero mean and $\sigma^2_{s,r}$ variance, and $e_{s,r}[n]$ is the AWGN noise with zero mean and $N_{s,r}$ variance. It is assumed that the channels $h_{s,d}$ and $h_{s,r}$ remain constant over at least two consecutive time-intervals $n-1$ and $n$. 

From~\eqref{eq:diffencode} and~\eqref{eq:r1r2recdata}, we have
\begin{align}
\label{eq:r1r2recdata1}
y_{s,r}[n]=y_{s,r}[n-1]x[n]+e'_{s,r}[n],
\end{align}\vspace*{-0.7em}

\hspace*{-0.95em}where $e'_{s,r}[n]=e_{s,r}[n]-e_{s,r}[n-1]x[n]\sim\mathcal{CN}\left(0,2N_{s,r}\right)$ is the AWGN noise and $\mathcal{CN}\left(\mu,\eta\right)$ denotes the complex Normal distribution with $\mu$ mean and $\eta$ variance. It can be seen from~\eqref{eq:r1r2recdata1} that for given $y_{s,r}[n-1]$ and $x[n]$, $y_{s,r}[n]\sim\mathcal{CN}\left(y_{s,r}[n-1]x[n],2N_{s,r}\right)$. By maximizing the conditional p.d.f. of $y_{s,r}[n]$ given that $y_{s,r}[n-1]$ and $x[n]$ are known in the relay, the ML decoder in the relay can be obtained as
\begin{align}
\label{eq:qam_decoder_relay}
{{x}_r}[n]=\text{arg}\ms\underset{x\in\mathcal{A}}{\text{max}}\:\text{Re}\left\{y^*_{s,r}[n]y_{s,r}[n-1]{x}\right\},
\end{align}\vspace*{-0.7em}

\hspace*{-0.95em}
where ${{x}_r}[n]\in\mathcal{A}$ and $\left|{{x}_r}[n]\right|^2=1$.
In the second phase, the relay differentially encodes ${x}_r[n]$ into ${v}_r[n]$ by using~\eqref{eq:diffencode} and transmits it to the destination. 
\section{Optimal and Suboptimal Decoders in the Destination}
\label{sec:MLDDest}
In this section, we will derive the ML and PL decoders of unitary and non-unitary constellations in the differential DF cooperative system.
\subsection{ML Decoder of $M$-PSK Data in the Destination}
\label{sub:MLDDest}
The data received at the destination from the relay in the $n$-th time interval will be
\begin{align}
\label{eq:relaytodest}
y_{r,d}[n]=h_{r,d}{v}_r[n]+e_{r,d}[n],
\end{align}\vspace*{-0.7em}

\hspace*{-0.95em}where $h_{r,d}\sim\mathcal{CN}\left(0,\sigma^2_{r,d}\right)$ is the channel gain of the relay-destination link and $e_{r,d}[n]\sim\mathcal{CN}\left(0,N_{r,d}\right)$ is the AWGN noise. It is assumed that $h_{r,d}$ remains constant over at least two consecutive time intervals $n-1$ and $n$. From~\eqref{eq:diffencode} and~\eqref{eq:relaytodest}, we have
\begin{align}
\label{eq:relaytodest1}
y_{r,d}[n]=y_{r,d}[n-1]{x}_r[n]+e'_{r,d}[n],
\end{align}
where 
$e'_{r,d}[n]=e_{r,d}[n]-e_{r,d}[n-1]{x}_r[n]$ is the zero mean AWGN noise with $2N_{r,d}$ variance. Depending upon the erroneous demodulation of the data of the source in the relay, the conditional p.d.f. of $y_{r,d}[n]$ can be written as
\begin{align}
\label{eq:mixture}
p_{y_{r,d}[n]}&\left(y|y_{r,d}[n-1],x[n]\right)\nonumber\\
&=\left(1-\epsilon\right)p_{y_{r,d}[n]}\left(y|y_{r,d}[n-1],{x}_r[n]=x[n]\right)\nonumber\\
&+\epsilon p_{y_{r,d}[n]}\left(y|y_{r,d}[n-1],{x}_r[n]\neq x[n]\right),
\end{align}
where $\epsilon$ is the average probability of error of the source-relay link. It can be noticed from~\eqref{eq:relaytodest1} that $p_{y_{r,d}[n]}\left(y|y_{r,d}[n-1],{x}_r[n]=x[n]\right)\sim\mathcal{CN}\left(y_{r,d}[n-1]x[n],2N_{r,d}\right)$.
From~\eqref{eq:relaytodest1} and~\cite[Section~III]{trail05}, it can be seen that $p_{y_{r,d}[n]}\left(y|y_{r,d}[n-1],{x}_r[n]\neq x[n]\right)$ denotes the p.d.f. of a Gaussian mixture random variable. With this observation we have
\begin{align}
\label{eq:gauss}
&p_{y_{r,d}[n]}\left(y|y_{r,d}[n-1],{x}_r[n]\neq x[n]\right)\nonumber\\
&=\frac{1}{2\pi N_{r,d}(M-1)}\sum^M_{i=1, i\neq p}e^{-\frac{1}{2N_{r,d}}\left|y_{r,d}[n]-y_{r,d}[n-1]x_i\right|^2}, 
\end{align}\\
where it is assumed that $x[n]=x_p$, $p\in\left\{1,2,...,M\right\}$, $x_p\in \mathcal{A}$, is the symbol transmitted by the source. By noticing that $y_{s,d}[n]\sim\mathcal{CN}\left(y_{s,d}[n-1]x[n],2N_{s,d}\right)$ and owing to independence of $y_{s,d}[n]$ and $y_{r,d}[n]$, an ML decoder can be obtained by maximizing the joint conditional p.d.f. of $y_{s,d}[n]$ and $y_{r,d}[n]$ as follows:
\begin{align}
\label{eq:MLDDest}
&\hat{x}[n]=\text{arg}\ms\underset{x\in\mathcal{A}}{\text{max}}\:\displaystyle\left\{\frac{1}{N_{s,d}}\text{Re}\left\{y^*_{s,d}[n]y_{s,d}[n-1]x\right\}\right. \nonumber\\
&\left. +\text{ln}\left((1-\epsilon)e^{\frac{1}{N_{r,d}}\text{Re}\left\{y^*_{r,d}[n]y_{r,d}[n-1]x\right\}}\right.\right.\nonumber\\
&\left.\left.+\frac{\epsilon}{M-1}\sum^M_{i=1, x_i\neq x}e^{\frac{1}{N_{r,d}}\text{Re}\left\{y^*_{r,d}[n]y_{r,d}[n-1]x_i\right\}}\right)\right\}.
\end{align}\\
It can be seen from~\eqref{eq:MLDDest} that the proposed decoder requires the destination to possess knowledge of the average probability of error of the source-relay link $\epsilon$ which is a function of the average signal-to-noise ratio (SNR) of the source-relay link. The relay can estimate the average SNR of the source-relay link and feed forward it to the destination. The destination can calculate the value of $\epsilon$ by using the average SNR of the source-relay link and use it for the ML decoding. Since statistics of a channel vary slowly as compared to the instantaneous channel values, the destination requires less frequent updating of the average SNR value of the source-relay link. Further,~\eqref{eq:MLDDest} is applicable to the $M$-PSK constellation, whereas, the existing decoders of the differential cooperative system~\cite{chen06,zhao07} are applicable to binary signaling only. For $M=2$,~\eqref{eq:MLDDest} reduces to the existing ML decoder of the binary data~\cite{chen06,zhao07}. However, the proposed decoder is computationally complex. Therefore, it is desired to have a low complexity decoder of the DF based differential cooperative system, which can provide maximum possible diversity.
\subsection{PL Decoder of $M$-PSK Data in the Destination}
\label{sub:Pl}
Let $y_{s,d}[n]=0, \forall n$, i.e., direct link between the source and the destination is absent, then a log-likelihood ratio (LLR) based \emph{symbol-wise} decoder can be obtained from~\eqref{eq:mixture} and~\eqref{eq:gauss} as follows:
\begin{align}
\label{eq:LLRDDest1}
\Lambda^c_{p,q}=\ln\left(\frac{a+g_1}{a+g_2}\right),
\end{align}\\
where $p,q=1,2,...,M, x_p,x_q\in \mathcal{A}$, $x_p\neq x_q$, $a=\frac{\epsilon}{M-1}\sum^M_{i=1\atop i\neq p,q}\displaystyle e^{\frac{1}{N_{r,d}}\text{Re}\left\{y^*_{r,d}[n]y_{r,d}[n-1]x_i\right\}}$, $g_1=(1-\epsilon)e^{\frac{1}{N_{r,d}}\text{Re}\left\{y^*_{r,d}[n]y_{r,d}[n-1]x_p\right\}}+\frac{\epsilon}{M-1}e^{\frac{1}{N_{r,d}}\text{Re}\left\{y^*_{r,d}[n]y_{r,d}[n-1]x_q\right\}}$, and $g_2=(1-\epsilon)e^{\frac{1}{N_{r,d}}\text{Re}\left\{y^*_{r,d}[n]y_{r,d}[n-1]x_q\right\}}+\frac{\epsilon}{M-1}\\ \times e^{\frac{1}{N_{r,d}}\text{Re}\left\{y^*_{r,d}[n]y_{r,d}[n-1]x_p\right\}}$. The LLR decoder of~\eqref{eq:LLRDDest1} is used as follows to decide about $x_p$ or $x_q$:
$\Lambda^c_{p,q} \overset{x_p}{\underset{x_q}{\gtrless}} 0$. 
Since $a,g_1,g_2\geq 0$, it can be shown after some simple algebra that the decision rule $\ln\left(\frac{a+g_1}{a+g_2}\right)\overset{x_p}{\underset{x_q}{\gtrless}} 0$ is the same as $\ln\left(\frac{g_1}{g_2}\right)\overset{x_p}{\underset{x_q}{\gtrless}} 0$. 
\begin{table}[!t]
\mycaption{Values of $T$ for different values of $\epsilon$ used in Fig.~\ref{fig:pl}.}
\begin{center}\small
\begin{tabular}{|c|c|c|c|c|c|c|}\hline 
\!\!\!\!$\epsilon$\!\!\! &$10^{-1}$& $10^{-2}$& $10^{-3}$ & $10^{-4}$ & $10^{-5}$ & $10^{-6}$\\ \hline
\!\!\!\!$T$\!\!\!&\!\!$\pm$\!\! 4.9053\!\!\!\! &\!\!$\pm$\!7.3032\!\!\!\!&\!\!$\pm$\!\! 9.6148\!\!\!\!&\!\!$\pm$ \!\!11.9183\!\!\!\!&\!\!$\pm$\!\! 14.2210\!\!\!\!&\!\! $\pm$\!\! 16.5236\!\!\!\!\\ \hline
\end{tabular}
\end{center}
\vspace*{-2em}
\label{tab:tb1}
\end{table}
\begin{figure}[t!]
  \begin{center}\hspace*{-1.05em}
    \psfig{figure=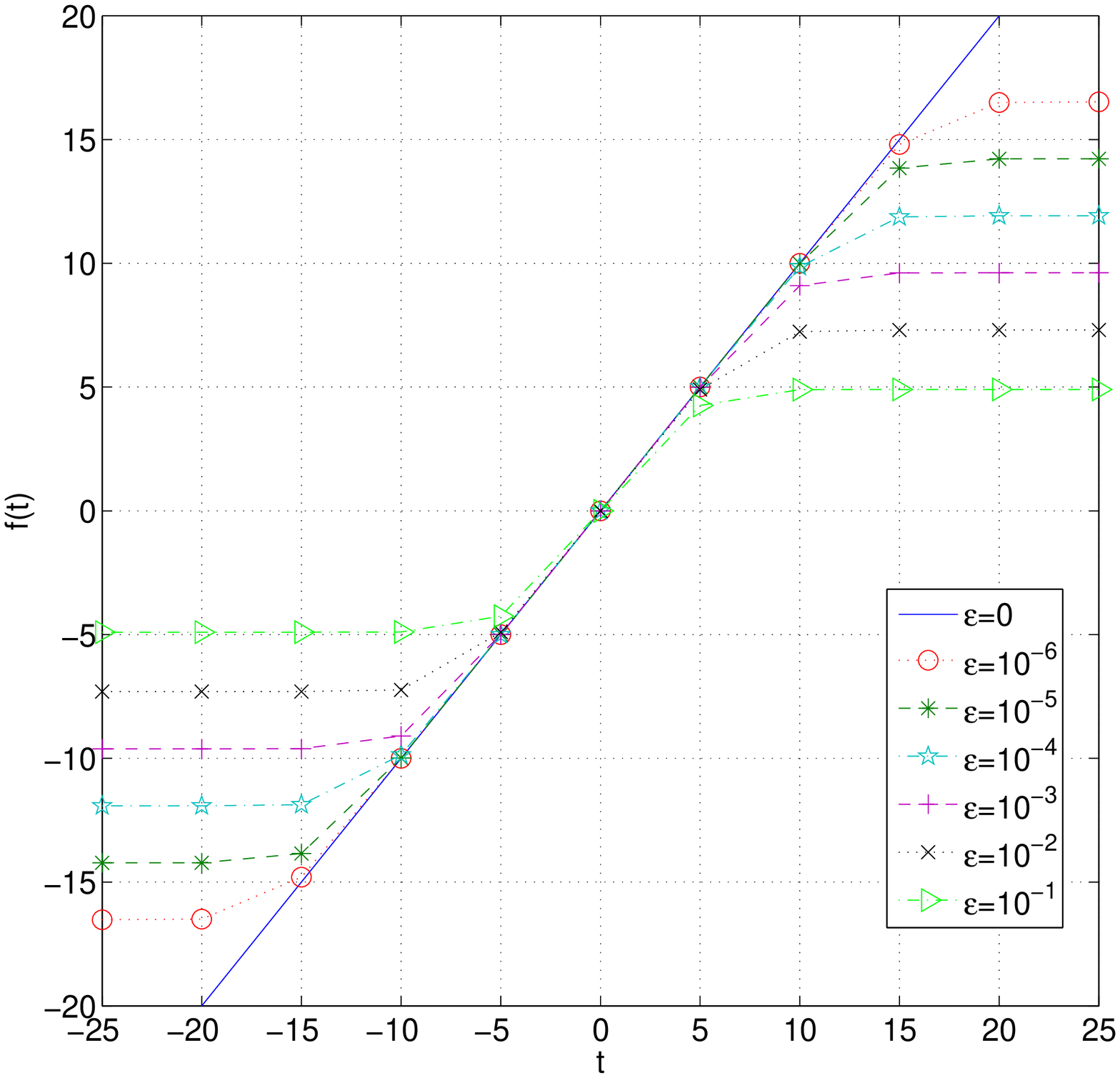,height=2.5in,width=3.75in}
    \vspace*{-1em}
    \mycaption{Plots of $f(t)$ versus $t$ for different values of $\epsilon$ and 16-PSK constellation.}
    \label{fig:pl}
    \vspace*{-0.5em}
  \end{center}
\end{figure}
Hence, we have the following LLR decoder from~\eqref{eq:LLRDDest1}:
\begin{align}
\label{eq:function}
\Lambda^c_{p,q}=f(t)=\text{ln}\left(\frac{(1-\epsilon)e^{t}+\frac{\epsilon}{M-1}}{(1-\epsilon)+\frac{\epsilon}{M-1}e^{t}}\right),
\end{align}\\
where 
$t=\frac{1}{N_{r,d}}\text{Re}\left\{y^*_{r,d}[n]y_{r,d}[n-1]\left(x_p-x_q\right)\right\}$.
It can be seen from~\eqref{eq:function} that $f(t)$ clipps to $T=\pm \text{ln}\left((M-1)(1-\epsilon)/\epsilon\right)$ for very large and very small values of $t$. Further, it is shown in Table~\ref{tab:tb1} and Fig.~\ref{fig:pl} for 16-PSK constellation that we can approximate $f(t)$ by a piecewise linear function as follows:
\begin{align}
\label{eq:PLfunc}
f(t)\approx f_{\text{PL}}(t)\triangleq \left\{\begin{array}{ccc}-T,&\text{if}& t< -T,\\t, &\text{if}&  -T\leq t \leq T,\\ T, &\text{if}&  t> T. \end{array} \right.
\end{align}\\
From~\eqref{eq:MLDDest} and~\eqref{eq:PLfunc}, we get the following \emph{suboptimal} PL decoder in the destination if the direct link between the source and destination is present:
\begin{align}
\label{eq:LLRDDest4}
&\Lambda_{p,q}\approx t_0
+f_{\text{PL}}(t),
\end{align}\\
where $t_0=\frac{1}{N_{s,d}}\text{Re}\left\{y^*_{s,d}[n]y_{s,d}[n-1]\left(x_p-x_q\right)\right\}$. The proposed PL decoder is applied in a pair-wise manner to the constellation points for taking a decision of the symbol transmitted by the source. For example, for QPSK constellation containing the following four signal points: $x_1,x_2,x_3,x_4$, the proposed PL decoder decides that $x_k, k=1,2,3,4,$ is the transmitted symbol if $\Lambda_{k,l}>0,\forall l\in\left\{1,2,3,4\right\}, k\neq l$, where $\Lambda_{k,l}$ is calculated from~\eqref{eq:LLRDDest4}.
 
\begin{figure}[t!]\vspace*{-1.0em}
  \begin{center}\hspace*{-0.0em}
    \psfig{figure=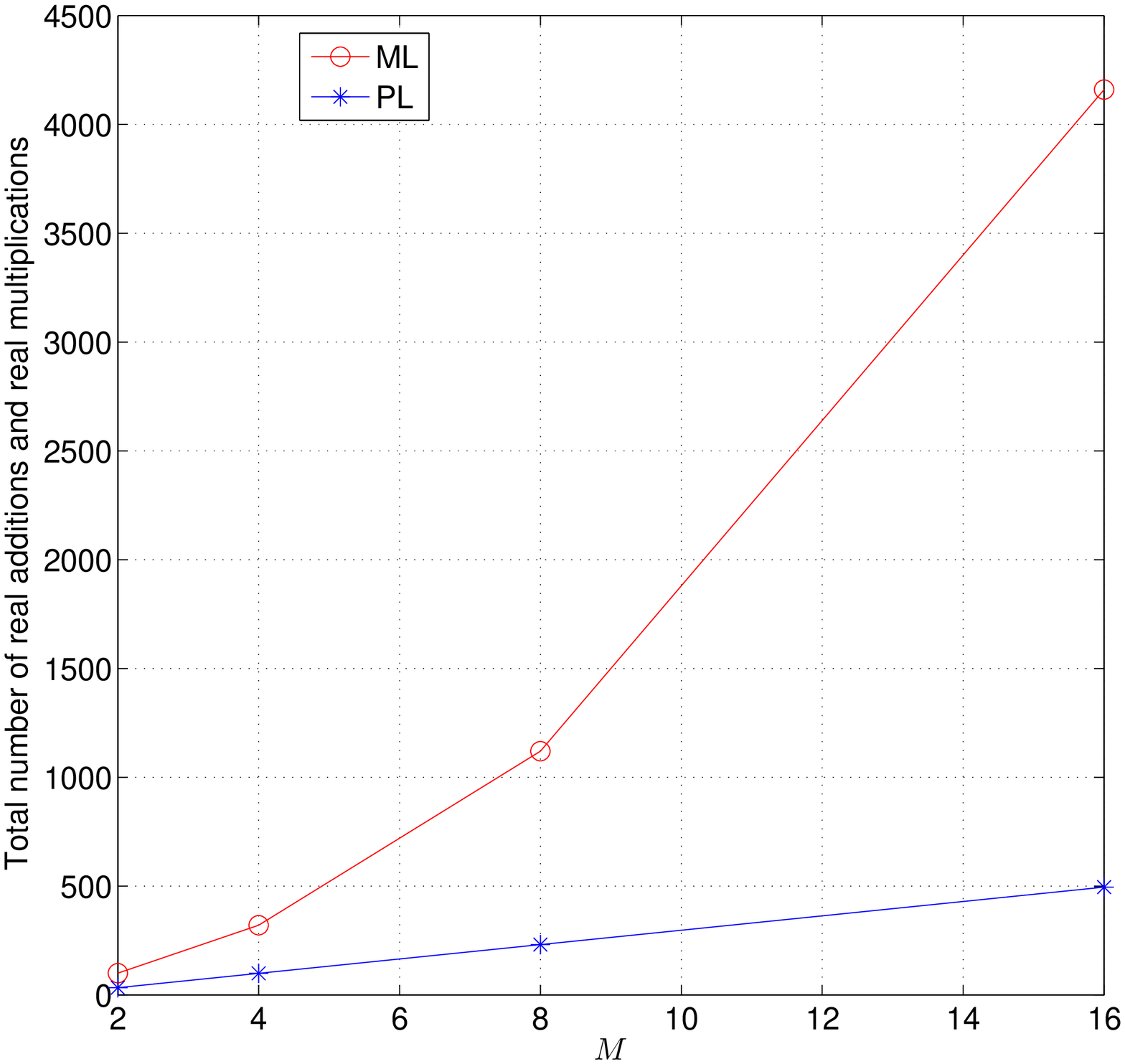,height=2.25in,width=3.5in}
    \vspace*{-1em}
    \mycaption{Plots of the total number of real additions and real multiplications required by the proposed ML and PL decoders for different $M$-PSK constellations.}
    \label{fig:deccomp}
    \vspace*{-1.0em}
  \end{center}
\end{figure}

The total number of the real additions and real multiplications required by the proposed ML decoder~\eqref{eq:MLDDest} and the proposed PL decoder~\eqref{eq:LLRDDest4} for decoding a symbol belonging to the $M$-PSK, $M\geq 2$, constellation can be obtained after some manipulations as $15M^2+20M$ and $33(M-1)$, respectively. We have plotted the decoding complexity (the total number of the real additions and real multiplications) of the proposed ML and PL decoders for different $M$-PSK constellations in Fig.~\ref{fig:deccomp}. It can be seen from Fig.~\ref{fig:deccomp} that the decoding complexity of the proposed PL decoder increases linearly, whereas, the decoding complexity of the proposed ML decoder increases exponentially with the size of the $M$-PSK constellation. Moreover, the proposed PL decoder significantly reduces the computational effort as compared to the proposed ML decoder for $M\geq 4$.
\subsection{ML and PL Decoders for Non-Unitary Constellations}
\label{sub:nonunitary}
The non-unitary constellations like $M$-QAM are useful for increasing the data rate and coding gain as compared to the $M$-PSK constellation in communication systems. Therefore, treatment of the differential modulation based cooperative communication system with non-unitary constellation is imperative. For non-unitary constellation like $M$-QAM, $|x[n]|^2$ is not necessarily equal to unity, therefore, we need to modify~\eqref{eq:diffencode} to satisfy the average power constraint as 
\begin{align}
\label{eq:diffencodeqam}
v[n]=v[n-1]\frac{x[n]}{\left|x[n-1]\right|},
\end{align}\\
where $\left|\cdot\right|$ denotes the absolute value. 
From~\eqref{eq:r1r2recdata} and~\eqref{eq:diffencodeqam}, we have 
\begin{align}
\label{eq:r1r2recdata2}
y_{s,r}[n]=y_{s,r}[n-1]\frac{x[n]}{\left|x[n-1]\right|}+e'_{s,r}[n],
\end{align}\\
where $e'_{s,r}[n]=e_{s,r}[n]-e_{s,r}[n-1]\frac{x[n]}{\left|x[n-1]\right|}\sim\mathcal{CN}\left(0,(1+\frac{\left|x[n]\right|^2}{\left|x[n-1]\right|^2})N_{s,r}\right)$.
As $y_{s,r}[n]\!\!\sim\!\mathcal{CN}\!\left(\!y_{s,r}[n-1]\frac{x[n]}{\left|x[n-1]\right|},(1\!+\!\frac{\left|x[n]\right|^2}{\left|x[n-1]\right|^2})N_{s,r}\!\right)$, therefore, 
the  relay demodulates $x[n]$ as
\begin{align}
\label{eq:qam_decoder_relay1}
&{x}_r[n]=\text{arg}\ms\underset{x\in\mathcal{\chi}}{\text{min}}\left\{\text{ln}\left(1+\frac{|x|^2}{\left|x[n-1]\right|^2}\right)+\frac{1/N_{s,r}}{1+\frac{|x|^2}{\left|x[n-1]\right|^2}}\right.\nonumber\\
&\left.\hspace*{5em}\times\left|y_{s,r}[n]-y_{s,r}[n-1]\frac{x}{\left|x[n-1]\right|}\right|^2\right\},
\end{align}\\
where $\mathcal{\chi}$ is an $M$-QAM constellation. 
It can be seen from~\eqref{eq:qam_decoder_relay1} that the differential decoder assumes that $x[n-1]$ is perfectly known to the receiver. Since it is not possible for the receiver to have perfect knowledge of $x[n-1]$, the differential decoder of~\eqref{eq:qam_decoder_relay1} can utilize an estimate of $x[n-1]$. %

From~\eqref{eq:destinationdata},~\eqref{eq:relaytodest}, and~\eqref{eq:diffencodeqam}, we can write
\begin{align}
\label{eq:dataqam}
y_{s,d}[n]&=y_{s,d}[n-1]\frac{x[n]}{\left|x[n-1]\right|}+e'_{s,d}[n],\nonumber\\
y_{r,d}[n]&=y_{r,d}[n-1]\frac{{x}_r[n]}{\left|x_r[n-1]\right|}+e'_{r,d}[n],
\end{align}\vspace*{-0.7em}

\hspace*{-1.0em}where $e'_{s,d}[n]=e_{s,d}[n]-e_{s,d}[n-1]\frac{x[n]}{\left|x[n-1]\right|}\sim\mathcal{CN}\left(0,(1+|x[n]|^2/\left|x[n-1]\right|^2)N_{s,d}\right)$ and $e'_{r,d}[n]=e_{r,d}[n]-e_{r,d}[n-1]\frac{{x}_r[n]}{\left|x_r[n-1]\right|}$ $\sim\mathcal{CN}\left(0,(1+|{x}_r[n]|^2/\left|x_r[n-1]\right|^2)N_{r,d}\right)$.
The joint conditional p.d.f. of $y_{s,d}[n]$ and $y_{r,d}[n]$ can be maximized to obtain the following ML decoder of the differential $M$-QAM data:
\begin{align}
\label{eq:MLDDestqam}
&\hat{x}[n]=\text{arg}\ms\underset{x\in\chi}{\text{max}}\:\left\{-\text{ln}\left(1+\frac{|x|^2}{\left|x[n-1]\right|^2}\right)\right. \nonumber\\
&\left.-\frac{\left|y_{s,d}[n]-y_{s,d}[n-1]\frac{x}{\left|x[n-1]\right|}\right|^2}{\left(1+\frac{|x|^2}{\left|x[n-1]\right|^2}\right)N_{s,d}}\right. \nonumber\\
&\left. +\text{ln}\left(\frac{(1-\epsilon)}{1+\frac{|x|^2}{\left|{x}_r[n-1]\right|^2}}e^{-\frac{\left|y_{r,d}[n]-y_{r,d}[n-1]\frac{x}{\left|{x}_r[n-1]\right|}\right|^2}{\left(1+\frac{|x|^2}{\left|{x}_r[n-1]\right|^2}\right)N_{r,d}}}+\frac{\epsilon}{M-1}\right.\right.\nonumber\\
&\left.\left.\times\sum^M_{i=1\atop x_i\neq x}\frac{1}{1+\frac{|x_i|^2}{\left|{x}_r[n-1]\right|^2}}e^{-\frac{\left|y_{r,d}[n]-y_{r,d}[n-1]\frac{x_i}{\left|{x}_r[n-1]\right|}\right|^2}{\left(1+\frac{|x_i|^2}{\left|{x}_r[n-1]\right|^2}\right)N_{r,d}}}\right)\right\}.
\end{align}\\
Since the destination does not have perfect knowledge of $x[n-1]$ and ${x}_r[n-1]$, it can utilize the estimated values of $x[n-1]$ and ${x}_r[n-1]$ in~\eqref{eq:MLDDestqam} for decoding the currently transmitted data of the source. The estimate of ${x}_r[n-1]$ in the destination can be obtained by using the following differential decoder:
\begin{align}
\label{eq:qam_decoder_relay_dest}
\hat{{x}}_r[n-1]=\text{arg}\!\!\ms\underset{x\in\mathcal{\chi}}{\text{min}}&\!\left\{\frac{\left|y_{r,d}[n-1]-y_{r,d}[n-2]\frac{x}{\left|{x}_r[n-2]\right|}\right|^2}{\left(1+\frac{|x|^2}{\left|{x}_r[n-2]\right|^2}\right)N_{r,d}}\right.\nonumber\\
&\left.+\text{ln}\left(\!1\!+\!\frac{|x|^2}{\left|{x}_r[n-2]\right|^2}\right)\!\!\right\}.
\end{align}\\ 
It will be shown in Subsection~\ref{sub:perdf} through Fig.~\ref{fig:mlplqam} that use of the estimated value of $x[n-1]$ in the relays and estimated values of $x[n-1]$ and ${x}_r[n-1]$ in the destination does not lead to error propagation in the performance of the destination receiver for different $M$-QAM constellations. 

After some algebra, the PL decoder of the differential $M$-QAM data can be obtained as follows:
\begin{align}
\label{eq:PL_qam}
&\Lambda_{p,q}\approx u_0
+f_{\text{PL}}(u_r),
\end{align}
where 
\begin{align}
\label{eq:v0vmqam}
u_0=&\ln\left(\frac{\left|x[n-1]\right|^2+\left|x_q\right|^2}{\left|x[n-1]\right|^2+\left|x_p\right|^2}\right)\nonumber\\
&\hspace*{0em}+\frac{\left|y_{s,d}[n]-y_{s,d}[n-1]\frac{x_q}{\left|x[n-1]\right|}\right|^2}{\left(1+\left|x_q\right|^2/\left|x[n-1]\right|^2\right)N_{s,d}}\nonumber\\
&\hspace*{0em}-\frac{\left|y_{s,d}[n]-y_{s,d}[n-1]\frac{x_p}{\left|x[n-1]\right|}\right|^2}{\left(1+\left|x_p\right|^2/\left|x[n-1]\right|^2\right)N_{s,d}},\nonumber\\
u_r=&\ln\left(\frac{\left|{x}_r[n-1]\right|^2+\left|x_q\right|^2}{\left|{x}_r[n-1]\right|^2+\left|x_p\right|^2}\right)\nonumber\\
&+\frac{\left|y_{r,d}[n]-y_{r,d}[n-1]\frac{x_q}{\left|{x}_r[n-1]\right|}\right|^2}{\left(1+{|x_q|^2}/{\left|{x}_r[n-1]\right|^2}\right)N_{r,d}}\nonumber\\
&-\frac{\left|y_{r,d}[n]-y_{r,d}[n-1]\frac{x_p}{\left|{x}_r[n-1]\right|}\right|^2}{\left(1+{|x_p|^2}/{\left|{x}_r[n-1]\right|^2}\right)N_{r,d}},
\end{align}\\
and $f_{\text{PL}}(u_r)$ is given in~\eqref{eq:PLfunc}. The proposed PL decoder of the $M$-QAM constellation given in~\eqref{eq:PL_qam} provides decoding complexity which is linear in $M$. Therefore, the proposed PL decoder of the $M$-QAM constellation also reduces the decoding complexity as compared to the proposed ML decoder~\eqref{eq:MLDDestqam}.\vspace*{0.5em}

\hspace*{-0.95em}\textbf{Remark:} \emph{We utilize one-way relaying in this paper for simplicity. The ML and PL decoders obtained in this paper can be extended to the two-way relaying based differential DF system. However, the extension can be more involved.} 
\section{Performance Analysis of the Differential Cooperative System with $M$-PSK Constellation}
\label{sec:performance}
In this section, we will analyze the uncoded symbol error rate of the proposed PL decoder of the DF based differential cooperative communication system with $M$-PSK constellation. 
\subsection{Average SER of the Differential DF System with $M$-PSK Constellation}
\label{sub:serPL}
Let $x[n]=x_p$, $x_p\in\mathcal{A}$, be the $M$-PSK symbol transmitted by the source and the destination wrongly decides that $\hat{x}[n]=x_q$, $x_q\in\mathcal{A}$, is the transmitted $M$-PSK symbol. The probability of error of the destination receiver can be expressed in terms of three mutually exclusive events.
The conditional uncoded pairwise error probability (PEP), given that the channel gains of the source-destination and relay-destination links are known in the destination, is
\begin{align}
\label{eq:condSER}
&P^{x_p,x_q}_e(h_{s,d},h_{r,d})\nonumber\\
&=\text{Pr}\left\{t_0-T<0|t<-T,x[n]\!\!=\!\!x_p\right\}\text{Pr}\left\{t<-T|x[n]\!\!=\!\!x_p\right\}\nonumber\\
&+\text{Pr}\left\{t_0+T<0|t>T,x[n]=x_p\right\}\text{Pr}\left\{t>T|x[n]=x_p\right\}\nonumber\\
&+\text{Pr}\left\{t_0+t<0,-T\leq t\leq T|x[n]=x_p\right\}, 
\end{align}
where $\text{Pr}\left\{\cdot\right\}$ represents the probability of an event. Let $\gamma_{s,d}=|h_{s,d}|^2/N_{s,d}$ and $\gamma_{r,d}=|h_{r,d}|^2/N_{r,d}$ be the instantaneous SNR of the source-destination and relay-destination link, respectively. 
It is shown in Appendix~\ref{app:condPEP} that the conditional uncoded PEP of the PL decoder in a differential cooperative system utilizing the DF protocol when $h_{s,d}$ and $h_{r,d}$ are known in the destination, will be
\begin{align}
\label{eq:condPEP}
&P^{x_p,x_q}_e(h_{s,d},h_{r,d})=P_{e_1}(h_{s,d})\left(P_{e_2}(h_{r,d})+P_{e_3}(h_{r,d})\right)\nonumber\\
&+P_{e_4}(h_{s,d})\left(P_{e_5}(h_{r,d})+P_{e_6}(h_{r,d})\right)+P_{e_7}(h_{s,d},h_{r,d})\nonumber\\
&+P_{e_8}(h_{s,d},h_{r,d})+P_{e_9}(h_{s,d},h_{r,d})+P_{e_{10}}(h_{s,d},h_{r,d}),
\end{align}
where the terms $P_{e_i}\left(\cdot\right), i=1,2,..,6$ and $P_{e_j}\left(\cdot,\cdot\right), j=7,..,10$ are given in Appendix~\ref{app:condPEP}.

If $h_{s,d}$ and $h_{r,d}$ are the complex-valued Gaussian random variables, then $\gamma_{s,d}$ and $\gamma_{r,d}$ will be Xi-square distributed with the following p.d.f.s~\cite{proak01}:
\begin{align}
\label{eq:Xisqp.d.f.} 
p_{\gamma_{s,d}}(\gamma)&=\frac{1}{\bar{\gamma}_{s,d}}e^{-\frac{\gamma}{\bar{\gamma}_{s,d}}}\nonumber\\
p_{\gamma_{r,d}}(\gamma)&=\frac{1}{\bar{\gamma}_{r,d}}e^{-\frac{\gamma}{\bar{\gamma}_{r,d}}},
\end{align}\vspace*{-0.7em}

\hspace*{-0.95em}
where $\bar{\gamma}_{s,d}=\sigma^2_{s,d}/N_{r,d}$ and $\bar{\gamma}_{r,d}=\sigma^2_{r,d}/N_{r,d}$ are the average SNRs of the source-destination and relay-destination links, respectively. The average uncoded PEP of decoding $x_q$ in place of $x_p$ can be obtained as
\begin{align}
\label{eq:avgser}
&P^{x_p,x_q}_e=E\left\{P^{x_p,x_q}_e(h_{s,d},h_{r,d})\right\}\nonumber\\
&=\int^{\infty}_0\int^\infty_0P^{x_p,x_q}_e(\gamma_1,\gamma_2)p_{\gamma_{s,d}}(\gamma_1)p_{\gamma_{r,d}}(\gamma_2)d\gamma_1\:d\gamma_2,
\end{align}\\
where $E\left\{\cdot\right\}$ denotes the expectation. 
It is shown in Appendix~\ref{app:condPEP} that the closed-form average PEP of decoding $x_q$ in place of $x_p$ in the DF based differential cooperative system using the $M$-PSK constellation will be 
\begin{align}
\label{eq:avgser1}
P^{x_p,x_q}_e=&P_{e_1}(P_{e_2}+P_{e_3})+P_{e_4}(P_{e_5}+P_{e_6})+P_{e_7}+P_{e_8}\nonumber\\
&+P_{e_9}+P_{e_{10}},
\end{align}
where the terms $P_{e_l},l=1,2,..,10$ are given in Appendix~\ref{app:condPEP}.

\begin{figure}[t!]
  \begin{center}\hspace*{0.0em}\vspace*{0.0em}
    \psfig{figure=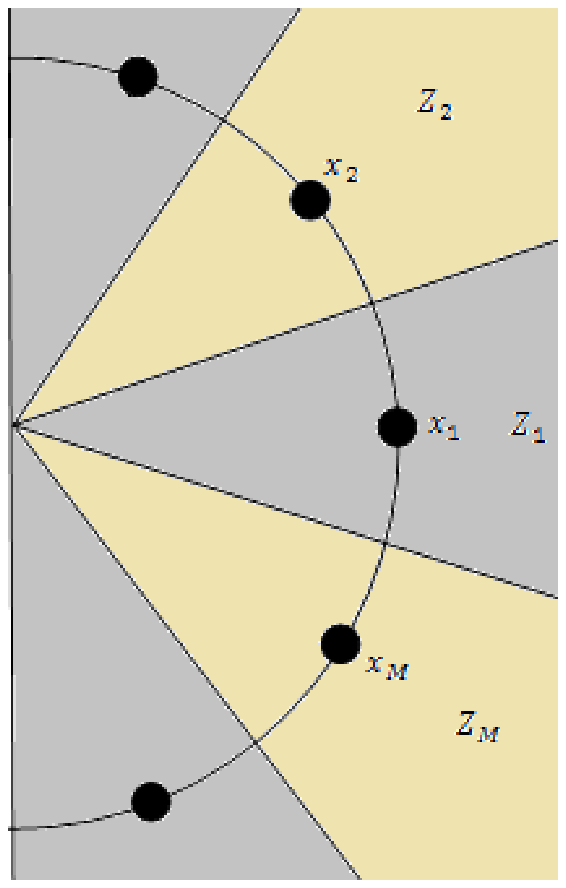,height=2.5in,width=1.8in}
    \vspace*{1.5em}
    \mycaption{Decision regions for $M$-PSK constellation.}
    \label{fig:const}
    \vspace*{0.0em}
  \end{center}
\end{figure}
The constellation diagram of an arbitrary $M$-PSK constellation is shown in Fig.\!~\ref{fig:const}. The decision boundaries corresponding to the symbol $x_1$ are also shown. From equiprobability of the constellation points it can be deduced that
\begin{align}
\label{eq:PSKSER}
\text{Pr}\left[\text{error}\right]=\frac{1}{M}\sum^M_{i=1}\text{Pr}\left[\text{error}|x_i\right]=\text{Pr}\left[\text{error}|x_1\right].
\end{align} 
Let $x_p=x_1=1$ is the transmitted symbol, then the SER of the $M$-PSK constellation can be approximated by using the nearest neighbors approach~\cite[Subsection~5.1.5]{golds05} as
\begin{align}
\label{eq:serqpsk3}
P_e\approx P^{x_2,1}_e+P^{x_{M},1}_e,
\end{align}
where $P^{x_2,1}_e$ and $P^{x_{M},1}_e$ are the average probabilities of error of decoding $x_2$ and $x_{M}$, respectively, in place of the originally transmitted symbol $x_1=1$. We can calculate $P^{x_2,1}_e$ and $P^{x_{M},1}_e$ by using~\eqref{eq:avgser1}. 
\subsection{Diversity Analysis of the Differential DF System with A Single Relay and $M$-PSK Constellation}
\label{sub:approxhgsnr}
Expression of the SER of the PL decoder given in~\eqref{eq:avgser1} and~\eqref{eq:serqpsk3} is very complicated and it is difficult to draw a conclusion from it. Therefore, we will derive an approximate SER of the proposed PL decoder with $M$-PSK constellation by ignoring the higher order noise in this subsection. For simplicity, we assume that $x_q$ is the transmitted symbol and the destination wrongly decides that $x_p$ was transmitted. It can be noticed that because of symmetry of the $M$-PSK constellation $P^{x_p,x_q}_e=P^{x_q,x_p}_e$.
By ignoring the higher order noise terms in $t_0$ and $t$, it can be shown after some manipulations that $t_0\sim \mathcal{N}\left(\text{Re}\left\{\gamma_{s,d}x^*_q\bar{x}\right\},\gamma_{s,d}\left|\bar{x}\right|^2\right)$ and $t\sim \mathcal{N}\left(\text{Re}\left\{\gamma_{r,d}{x}^*_r\bar{x}\right\},\gamma_{r,d}\left|\bar{x}\right|^2\right)$, where $\bar{x}=x_p-x_q$ and $x_r$ is the symbol transmitted by the relay. After many algebraic manipulations, the average pairwise error probability of the proposed PL decoder can be obtained by ignoring the higher order noise as
\begin{align}
\label{eq:pep}
P^{x_p,x_q}_e\approx I_1(\bar{\gamma},T)+I_2(\bar{\gamma},T)+I_3(\bar{\gamma},T)+I_4(\bar{\gamma},T),
\end{align}
where
\begin{align}
\label{eq:int1}
&I_1(\bar{\gamma},T)=(1-\epsilon)\frac{1}{\bar{\gamma}_{s,d}} \int^{\infty}_0 Q\left(\frac{T-  z_q\gamma_{s,d}}{\left|\bar{x}\right|\sqrt{\gamma_{s,d}}}\right)e^{-\frac{\gamma_{s,d}}{\bar{\gamma}_{s,d}}}d\gamma_{s,d} \nonumber\\
&\times\frac{1}{\bar{\gamma}_{r,d}} \int^{\infty}_0 Q\left(\frac{T+ z_q\gamma_{r,d}}{\left|\bar{x}\right|\sqrt{\gamma_{r,d}}}\right)e^{-\frac{\gamma_{r,d}}{\bar{\gamma}_{r,d}}}d\gamma_{r,d}\nonumber\\
&+\frac{\epsilon}{M-1} \frac{1}{\bar{\gamma}_{s,d}} \int^{\infty}_0 Q\left(\frac{T-  z_q\gamma_{s,d}}{\left|\bar{x}\right|\sqrt{\gamma_{s,d}}}\right)e^{-\frac{\gamma_{s,d}}{\bar{\gamma}_{s,d}}}d\gamma_{s,d} \nonumber\\
&\times\sum^M_{i=1\atop i\neq q}\left(\frac{1}{\bar{\gamma}_{r,d}} \int^{\infty}_0 Q\left(\frac{T+ z_i\gamma_{r,d}}{\left|\bar{x}\right|\sqrt{\gamma_{r,d}}}\right)e^{-\frac{\gamma_{r,d}}{\bar{\gamma}_{r,d}}}d\gamma_{r,d}\right),\nonumber\\
&I_2(\bar{\gamma},T)=(1-\epsilon)\frac{1}{\bar{\gamma}_{s,d}} \int^{\infty}_0 Q\left(\frac{-T-  z_q\gamma_{s,d}}{\left|\bar{x}\right|\sqrt{\gamma_{s,d}}}\right)e^{-\frac{\gamma_{s,d}}{\bar{\gamma}_{s,d}}}d\gamma_{s,d}\nonumber\\
&\times \frac{1}{\bar{\gamma}_{r,d}} \int^{\infty}_0 Q\left(\frac{T- z_q\gamma_{r,d}}{\left|\bar{x}\right|\sqrt{\gamma_{r,d}}}\right)e^{-\frac{\gamma_{r,d}}{\bar{\gamma}_{r,d}}}d\gamma_{r,d}\nonumber\\
&+\frac{\epsilon}{M-1} \frac{1}{\bar{\gamma}_{s,d}} \int^{\infty}_0 Q\left(\frac{-T-  z_q\gamma_{s,d}}{\left|\bar{x}\right|\sqrt{\gamma_{s,d}}}\right)e^{-\frac{\gamma_{s,d}}{\bar{\gamma}_{s,d}}}d\gamma_{s,d} \nonumber\\
&\times\sum^M_{i=1\atop i\neq q}\left(\frac{1}{\bar{\gamma}_{r,d}} \int^{\infty}_0 Q\left(\frac{T- z_i\gamma_{r,d}}{\left|\bar{x}\right|\sqrt{\gamma_{r,d}}}\right)e^{-\frac{\gamma_{r,d}}{\bar{\gamma}_{r,d}}}d\gamma_{r,d}\right),\nonumber\\
%
&I_3(\bar{\gamma},T)=\frac{1-\epsilon}{\bar{\gamma}_{s,d}\bar{\gamma}_{r,d}}\int^\infty_{\gamma_{s,d}=0}\int^\infty_{\gamma_{r,d}=0}\frac{1}{\sqrt{2\pi\gamma_{r,d}}\left|\bar{x}\right|}\nonumber\\
&\times\int^{T}_{-T} e^{-\frac{1}{2}\frac{\left(w-z_q\gamma_{r,d}\right)^2}{\left|\bar{x}\right|^2{\gamma_{r,d}}}} Q\left(\frac{-w-z_q\gamma_{s,d}}{\left|\bar{x}\right|\sqrt{\gamma_{s,d}}}\right)e^{-\frac{\gamma_{s,d}}{\bar{\gamma}_{s,d}}} \nonumber\\
&\times e^{-\frac{\gamma_{r,d}}{\bar{\gamma}_{r,d}}} dw d\gamma_{r,d} d\gamma_{s,d},\nonumber\\
&I_4(\bar{\gamma},T)=\frac{\epsilon}{\bar{\gamma}_{s,d}\bar{\gamma}_{r,d}}\int^\infty_{\gamma_{s,d}=0}\int^\infty_{\gamma_{r,d}=0}\frac{1}{\sqrt{2\pi\gamma_{r,d}}\left|\bar{x}\right|(M-1)}\nonumber\\
&\times\int^{T}_{-T}Q\left(\frac{-w-z_q\gamma_{s,d}}{\left|\bar{x}\right|\sqrt{\gamma_{s,d}}}\right)\sum^{M}_{i=1 \atop i\neq q}e^{-\frac{1}{2}\frac{\left(w-z_i\gamma_{r,d}\right)^2}{\left|\bar{x}\right|^2{\gamma_{r,d}}}}e^{-\frac{\gamma_{s,d}}{\bar{\gamma}_{s,d}}}\nonumber\\
&\times e^{-\frac{\gamma_{r,d}}{\bar{\gamma}_{r,d}}}dw d\gamma_{r,d} d\gamma_{s,d},
\end{align}
where $Q\left(\cdot\right)$ is the Q-function~\cite[Eq.~(2.3.10)]{proak08}, $z_q=\text{Re}\left\{x_q\bar{x}^*\right\}$, and $z_i=\text{Re}\left\{x_i\bar{x}^*\right\}$. It is difficult to solve the integration in~\eqref{eq:pep} analytically, therefore, the approximate average SER of the PL decoder can be obtained numerically from~\eqref{eq:serqpsk3} and~\eqref{eq:pep}. We can use~\eqref{eq:pep} for analytically finding the diversity of the PL decoder. 

Let us assume that the average SNR of all links involved in the cooperation is equal and approaching to infinity with the same rate, i.e., $\bar{\gamma}_{s,d}=\bar{\gamma}_{r,d}=\bar{\gamma}_{s,r}=\bar{\gamma}\rightarrow\infty$, where $\bar{\gamma}_{s,r}$ is the average SNR of the source-relay link.
From~\cite[Section~5.1]{simon05}, it can be shown that for $M$-PSK constellation and large values of $\bar{\gamma}$, $\epsilon\propto \frac{1}{\bar{\gamma}}$. Therefore, for large values of $\bar{\gamma}$, $T\propto \pm {\ln}\bar{\gamma}$.  
\begin{lemma}
\label{lem:1}
The following three inequalities are valid for integrals of the Q-function:
\begin{align}
\label{eq:res4}
&\frac{1}{\bar{\gamma}} \int^{\infty}_0 Q\left(\frac{Z+  a_1\gamma}{\sqrt{b_1\gamma}}\right)e^{-\frac{\gamma}{\bar{\gamma}}}d\gamma\leq \frac{1}{\bar{\gamma}} \int^{\infty}_0 e^{-\frac{\left(Z+ a_1\gamma\right)^2}{2{b_1\gamma}}}e^{-\frac{\gamma}{\bar{\gamma}}}d\gamma,\nonumber\\
&\frac{1}{\bar{\gamma}} \int^{\infty}_0 Q\left(\frac{-Z+a_1\gamma}{\sqrt{b_1\gamma}}\right)e^{-\frac{\gamma}{\bar{\gamma}}}d\gamma \leq \sum^\infty_{k=1}(-1)^{k+1}\frac{(Z/\bar{\gamma})^k}{k!a^k_2}\nonumber\\
&\hspace*{10em}+\frac{1}{\bar{\gamma}} \int^{\infty}_0 e^{-\frac{\left(Z-a_1\gamma\right)^2}{2{b_1\gamma}}}e^{-\frac{\gamma}{\bar{\gamma}}}d\gamma,\nonumber\\
&\frac{1}{\bar{\gamma}} \int^{\infty}_0 Q\left(\frac{Z-  a_1\gamma}{\sqrt{b_1\gamma}}\right)e^{-\frac{\gamma}{\bar{\gamma}}}d\gamma\leq \eta\left({Z}/{\bar{\gamma}}\right)^{\alpha}\nonumber\\
&\hspace*{10em}+\frac{1}{\bar{\gamma}} \int^{\infty}_0 e^{-\frac{\left(Z-a_1\gamma\right)^2}{2{b_1\gamma}}}e^{-\frac{\gamma}{\bar{\gamma}}}d\gamma,
\end{align}\\
where $Z,a_1,b_1,\eta>0$, and $\alpha\geq 1$.
\end{lemma}
Lemma~\ref{lem:1} can be proved with help of the Chernov bound $Q(x)\leq e^{-x^2/2}, x\geq 0$~\cite[Section~4.1.1]{simon05}.
\begin{lemma}
\label{lem:2} 
If $Z=\ln\bar{\gamma}$, $a_2\neq 0$, $a_3$ is an arbitrary real-valued constant, and $b_2>0$, then 
\begin{align}
\label{eq:hgsnr1}
\underset{\bar{\gamma}\rightarrow \infty}{{\lim}}\frac{1}{\bar{\gamma}} \int^{\infty}_0 
e^{-\frac{\left(Z-a_2\gamma\right)^2}{b_2\gamma}}e^{-\frac{\gamma}{\bar{\gamma}}}d\gamma\approx \frac{c_1}{\bar{\gamma}},%
\end{align}
\begin{align}
\label{eq:res5}
\underset{\bar{\gamma}\rightarrow \infty}{{\lim}}\frac{1}{\bar{\gamma}} \int^{\infty}_0 \frac{1}{\sqrt{\gamma}}
e^{-\frac{\left(Z-a_3\gamma\right)^2}{b_2\gamma}}e^{-\frac{\gamma}{\bar{\gamma}}}d\gamma= \frac{c_2}{\bar{\gamma}},
\end{align}
and
\begin{align}
\label{eq:hgsnr2}
\underset{\bar{\gamma}\rightarrow \infty}{{\lim}}\frac{\epsilon}{\bar{\gamma}} \int^{\infty}_0 
e^{-\frac{Z^2}{b_2\gamma}}e^{-\frac{\gamma}{\bar{\gamma}}}d\gamma\approx \frac{c_3}{\bar{\gamma}},
\end{align}\\
where $c_1,c_2,c_3$ are positive constants. 
\end{lemma}\vspace*{0.25em}
Lemma~\ref{lem:2} can be proved by applying~\cite[Eq.~(3.471.9)]{grand00} in the left hand side of~\eqref{eq:hgsnr1},~\eqref{eq:res5}, and~\eqref{eq:hgsnr2}, then using the relations $K_{\nu}(x)\approx\sqrt{\pi/(2x)}e^{-x}, x>>0$ in~\eqref{eq:hgsnr1}, $K_{1/2}(x)=\sqrt{\pi/(2x)}e^{-x}$ in~\eqref{eq:res5}, and $\epsilon\approx\frac{c_3}{\bar{\gamma}}$ and $K_\nu(x)\approx\frac{1}{2} \Gamma(\nu)\left(\frac{1}{2}x\right)^{-\nu}$, $x\rightarrow 0$ in~\eqref{eq:hgsnr2}
, where $K_{\nu}(\cdot)$ is the modified Bessel function~\cite[Section~9.6.1]{abram72} and $\Gamma\left(\cdot\right)$ is the Gamma function~\cite[Eq.~(6.1.1)]{abram72}. 
We have also used the fact that $\ln\bar{\gamma}$ varies extremely slowly as compared to $\bar{\gamma}$ for $\bar{\gamma}\rightarrow\infty$. Therefore, for diversity related calculations $\ln\bar{\gamma}$ can be assumed constant relative to $\bar{\gamma}$. 

Let us consider the QPSK constellation for proving the diversity of the proposed PL decoder for simplicity. For QPSK constellation, $z_q=\text{Re}\left\{x_q\bar{x}^*\right\}$ is negative, and $z_i=\text{Re}\left\{x_i\bar{x}^*\right\}$ can be negative, zero, or positive. Therefore, by using Lemmas~\ref{lem:1} and~\ref{lem:2} in~\eqref{eq:int1} it can be shown that
\begin{align}
\label{eq:div12}
\underset{\bar{\gamma}\rightarrow \infty}{\text{lim}}I_1(\bar{\gamma},T)\propto \frac{c_4}{\bar{\gamma}^2}+{f_1}\left({\bar{\gamma}}\right),\nonumber\\
\underset{\bar{\gamma}\rightarrow \infty}{\text{lim}}I_2(\bar{\gamma},T)\propto \frac{c_5}{\bar{\gamma}^2}+{f_2}\left(\bar{\gamma}\right),
\end{align}\\
where $c_4,c_5$ are constants and ${f_1}\left({\bar{\gamma}}\right),{f_2}\left({\bar{\gamma}}\right)$ are functions containing summation terms each decaying at rate higher than ${\bar{\gamma}^{-2}}$ for $\bar{\gamma}\rightarrow \infty$.

By observing the fact that $Q(x)$ is a decaying function of $x$, we have
\begin{align}
\label{eq:third1}
I_3(\bar{\gamma},T)\leq &\frac{2T\left(1-\epsilon\right)}{\sqrt{2\pi}\left|\bar{x}\right|\bar{\gamma}^2}\int^\infty_0 Q\left(\frac{-T-z_q\gamma}{\left|\bar{x}\right|\sqrt{\gamma}}\right)e^{-\frac{\gamma}{\bar{\gamma}}}d\gamma\nonumber\\
&\times\int^\infty_{0}\frac{1}{\sqrt{\gamma}}e^{-\frac{1}{2}\frac{\left(-T-z_q\gamma\right)^2}{\left|\bar{x}\right|^2{\gamma}}} e^{-\frac{\gamma}{\bar{\gamma}}}d\gamma.
\end{align}\\
By using Lemmas~\ref{lem:1},~\ref{lem:2} 
in~\eqref{eq:third1}, 
it can be shown that
\begin{align}
\label{eq:i31}
\underset{\bar{\gamma}\rightarrow\infty}{\text{lim}}I_3(\bar{\gamma},T)\propto\frac{c_6}{\bar{\gamma}^2}+ {f_3}\left({\bar{\gamma}}\right),
\end{align}\\
where $c_6$ is a positive constant and ${f_3}\left({\bar{\gamma}}\right)$ is a function containing summation terms each decaying at rate higher than ${\bar{\gamma}^{-2}}$ for $\bar{\gamma}\rightarrow \infty$. Similarly, it can be shown that 
\begin{align}
\label{eq:i32}
\underset{\bar{\gamma}\rightarrow\infty}{\text{lim}}I_4(\bar{\gamma},T)\propto
{f_4}\left({\bar{\gamma}}\right),
\end{align}\\
where 
${f_4}\left({\bar{\gamma}}\right)$ is a function containing summation terms each decaying at rate higher than ${\bar{\gamma}^{-2}}$ for $\bar{\gamma}\rightarrow \infty$.

It can be seen from~\eqref{eq:pep},~\eqref{eq:int1},~\eqref{eq:div12},~\eqref{eq:i31}, and~\eqref{eq:i32}, that the average probability of decoding $x_q$ as $x_p$ by a PL decoder decays as $\bar{\gamma}^{-2}$ at $\bar{\gamma}\rightarrow \infty$ and, therefore, the proposed PL decoder achieves a second order diversity.
\section{Differential DF System with Multiple Relays}
\label{sec:mult}
Let us consider a general case of $N$, $N\geq 1$, relays cooperating along with the direct transmission. It is assumed that the relays and the source use $N+1$ time intervals, in order to transmit the data \emph{orthogonally} to the destination. 
\subsection{ML and PL Decoders for $M$-QAM Constellation}
It can be shown after some algebra that for the differential $M$-QAM data and $N$ relays the ML decoder will be
\begin{align}
\label{eq:MLDDestqam1}
&\hat{x}[n]=\text{arg}\ms\underset{x\in\chi}{\text{max}}\:\left\{-\frac{\left|y_{s,d}[n]-y_{s,d}[n-1]\frac{x}{\left|x[n-1]\right|}\right|^2}{\left(1+\frac{|x|^2}{\left|x[n-1]\right|^2}\right)N_{s,d}}\right. \nonumber\\
&\left. +\sum^N_{m=1}\text{ln}\left(\frac{(1-\epsilon_m)}{1+\frac{|x|^2}{\left|{x}_m[n-1]\right|^2}}e^{-\frac{\left|y_{r_m,d}[n]-y_{r_m,d}[n-1]\frac{x}{\left|{x}_m[n-1]\right|}\right|^2}{\left(1+\frac{|x|^2}{\left|{x}_m[n-1]\right|^2}\right)N_{r_m,d}}}\right.\right.\nonumber\\
&\left.\left.+\frac{\epsilon_m}{M-1}\!\!\!\sum^M_{i=1\atop x_i\neq x}\!\!\!\frac{1}{1+\frac{|x_i|^2}{\left|{x}_m[n-1]\right|^2}}e^{-\frac{\left|y_{r_m,d}[n]-y_{r_m,d}[n-1]\frac{x_i}{\left|{x}_m[n-1]\right|}\right|^2}{\left(1+\frac{|x_i|^2}{\left|{x}_m[n-1]\right|^2}\right)N_{r_m,d}}}\!\right)\right.\nonumber\\
&\left.\hspace*{9em}-\text{ln}\left(1+\frac{|x|^2}{\left|x[n-1]\right|^2}\right)\right\},
\end{align}\\
where $x_m[n]$ is the demodulated symbol in the $m$-th, $m=1,...,N$, relay in the $n$-th time interval, $\epsilon_m$ is the average probability of error of the link between the source and the $m$-th relay, $y_{r_m,d}[n]$ is the signal received from the $m$-th relay by the destination in the $n$-th time interval, and $N_{r_m,d}$ is the variance of the AWGN noise of the link between the $m$-th relay and the destination. 
Since the destination does not have perfect knowledge of $x[n-1]$ and ${x}_m[n-1]$, it can utilize the estimated values of $x[n-1]$ and ${x}_m[n-1]$ in~\eqref{eq:MLDDestqam1} for decoding the currently transmitted data of the source. The estimate of ${x}_m[n-1]$ in the destination can be obtained by using the following differential decoder:
\begin{align}
\label{eq:qam_decoder_relay_dest1}
\hat{x}_m[n-1]\!\!=\!\!\text{arg}\!\!\ms\underset{x\in\mathcal{\chi}}{\text{min}}&\!\left\{\!\!\frac{\left|y_{r_m,d}[n-1]\!-\!y_{r_m,d}[n-2]\frac{x}{\left|x_m[n-2]\right|}\right|^2}{\left(1+\frac{|x|^2}{\left|x_m[n-2]\right|^2}\right)N_{r_m,d}}\right.\nonumber\\
&\left.+\text{ln}\left(\!1\!+\!\frac{|x|^2}{\left|x_m[n-2]\right|^2}\right)\!\!\right\}.
\end{align}\\
Moreover, the PL decoder for the differential $M$-QAM data and $N$ relays can be obtained after some algebra as
\begin{align}
\label{eq:PL_qam1}
&\Lambda_{p,q}\approx u_0
+\sum^N_{m=1}f_{\text{PL}}(u_m),
\end{align}
where 
\begin{align}
\label{eq:v0vmqam1}
u_0=&\ln\left(\frac{\left|x[n-1]\right|^2+\left|x_q\right|^2}{\left|x[n-1]\right|^2+\left|x_p\right|^2}\right)\nonumber\\
&+\frac{\left|y_{s,d}[n]-y_{s,d}[n-1]\frac{x_q}{\left|x[n-1]\right|}\right|^2}{\left(1+\left|x_q\right|^2/\left|x[n-1]\right|^2\right)N_{s,d}}\nonumber\\
&\hspace*{0em}-\frac{\left|y_{s,d}[n]-y_{s,d}[n-1]\frac{x_p}{\left|x[n-1]\right|}\right|^2}{\left(1+\left|x_p\right|^2/\left|x[n-1]\right|^2\right)N_{s,d}},\nonumber\\
u_m=&\ln\left(\frac{\left|x_m[n-1]\right|^2+\left|x_q\right|^2}{\left|x_m[n-1]\right|^2+\left|x_p\right|^2}\right)\nonumber\\
&+\frac{\left|y_{r_m,d}[n]-y_{r_m,d}[n-1]\frac{x_q}{\left|x_m[n-1]\right|}\right|^2}{\left(1+{|x_q|^2}/{\left|x_m[n-1]\right|^2}\right)N_{r_m,d}}\nonumber\\
&\hspace*{0em}-\frac{\left|y_{r_m,d}[n]-y_{r_m,d}[n-1]\frac{x_p}{\left|x_m[n-1]\right|}\right|^2}{\left(1+{|x_p|^2}/{\left|x_m[n-1]\right|^2}\right)N_{r_m,d}},
\end{align}
\begin{align}
\label{eq:PLfuncmult}
f_{\text{PL}}(u_m)\triangleq \left\{\begin{array}{ccc}-T_m,&\text{if}& u_m< -T_m,\\u_m, &\text{if}&  -T_m\leq u_m \leq T_m,\\ T_m, &\text{if}&  u_m> T_m, \end{array} \right.
\end{align}\\
and $T_m=\pm \text{ln}\left((M-1)(1-\epsilon_m)/\epsilon_m\right)$.
\subsection{ML and PL Decoders for $M$-PSK Constellation}
By substituting $\left|x\right|=\left|x[n]\right|=\left|x_m[n]\right|=1$ in~\eqref{eq:MLDDestqam1} and after some manipulations, the ML decoder for the $M$-PSK constellation with multiple relays can be obtained as 
\begin{align}
\label{eq:MLDDest1}
&\hat{x}[n]=\text{arg}\ms\underset{x\in\mathcal{A}}{\text{max}}\:\left\{\frac{1}{N_{s,d}}\text{Re}\left\{y^*_{s,d}[n]y_{s,d}[n-1]x\right\}\right. \nonumber\\
&\left. +\sum^N_{m=1}\text{ln}\left((1-\epsilon_m)e^{\frac{1}{N_{r_m,d}}\text{Re}\left\{y^*_{r_m,d}[n]y_{r_m,d}[n-1]x\right\}}\right.\right.\nonumber\\
&\left.\left.+\frac{\epsilon_m}{M-1}\sum^M_{i=1, x_i\neq x}e^{\frac{1}{N_{r_m,d}}\text{Re}\left\{y^*_{r_m,d}[n]y_{r_m,d}[n-1]x_i\right\}}\right)\right\}.
\end{align}\\
Similarly, the PL decoder for the $M$-PSK constellation can be obtained from~\eqref{eq:PL_qam1} as
\begin{align}
\label{eq:LLRDDest44}
&\Lambda_{p,q}\approx t_0
+\sum^N_{m=1}f_{\text{PL}}(t_m),
\end{align}\\
where $t_0=\frac{1}{N_{s,d}}\text{Re}\left\{y^*_{s,d}[n]y_{s,d}[n-1]\left(x_p-x_q\right)\right\}$,  $t_m=\frac{1}{N_{r_m,d}}\text{Re}\left\{y^*_{r_m,d}[n]y_{r_m,d}[n-1]\left(x_p-x_q\right)\right\}$, and $f_{\text{PL}}(t_m)$ is given in~\eqref{eq:PLfuncmult}.
\subsection{Asymptotic SER Analysis of the PL Decoder with Multiple Relays and $M$-PSK Constellation}
\label{sub:asyanal}
Let us assume asymptotically that $\bar{\gamma}_{s,r_m}\rightarrow \infty$ and $\bar{\gamma}_{s,d}=\bar{\gamma}_{r_m,d}=\bar{\gamma}$, where $\bar{\gamma}_{s,r_m}$ and $\bar{\gamma}_{r_m,d}$ are the average SNRs of the links between the source and the $m$-th relay and between the $m$-th relay and the destination, respectively. Since $\bar{\gamma}_{s,r_m}\rightarrow \infty$, $\epsilon_m\rightarrow 0$ and $T_m\rightarrow \infty$, i.e., the $m$-th relay becomes error-free under this asymptotic condition. Hence, the asymptotic SER of the proposed PL decoder with multiple relays can be obtained by assuming that all the relays are error-free. The conditional uncoded PEP of the proposed PL decoder~\eqref{eq:LLRDDest44} with $M$-PSK constellation and $N$ error-free relays will be
\begin{align}
\label{eq:condSERasyp}
&P^{x_p,x_q}_e(h_{s,d},h_{r_m,d})\nonumber\\
&=\text{Pr}\!\left\{\!t_0\!\!+\!\!\!\sum^N_{m=1}\!\!t_m\!<\!0,-\infty\!\leq\! t_m\!\leq \infty|x[n]\!\!=\!\!x_p, {x}_m[n]\!\!=\!\!x_p\!\right\}.
\end{align}\\
By using the results of quadratic forms in complex Gaussian variables~\cite{biyar93} in~\eqref{eq:condSERasyp}, we have 
\begin{align}
\label{eq:condPEPNrelays}
&P^{x_p,x_q}_e(h_{s,d},h_{r_m,d})\!\!=\!\!e^{\left(\frac{\beta}{4}-\frac{3|\bar{x}|^2}{2}\right)\gamma_t}\!\sum^\infty_{k=0}\!\!\sum^{k+N}_{n=0}\!\!\frac{(2|\bar{x}|^2-\beta)^k\gamma^k_t}{2^{N+n+k+1} k!}\nonumber\\
&\hspace*{9em}\times L^N_n\left(-\frac{(2|\bar{x}|^2+\beta)}{4}\gamma_t\right),
\end{align}\\
where $L^{\alpha}_n\left(\cdot\right)$ is the generalized Laguerre polynomial~\cite[pg.~775]{abram72}, $\beta=2\text{Re}\left\{x^*_p\bar{x}\right\}$, and $\gamma_t=\gamma_{s,d}+\sum^{N}_{m=1}\gamma_{r_m,d}$, where $\gamma_{r_m,d}$ is the instantaneous SNR of the link between the $m$-th relay and the destination. The distribution of $\gamma_t$ can be obtained from~\eqref{eq:Xisqp.d.f.} and~\cite[Eq.~(2.1.110)]{proak01} as
\begin{align}
\label{eq:Xisqpdfgen}
p_{\gamma_t}(\gamma)=\frac{\gamma^N}{\Gamma(N+1)\bar{\gamma}^{N+1}}e^{-\frac{\gamma}{\bar{\gamma}}}.
\end{align}\\
By using the series expansion of the generalized Laguerre polynomial~\cite[Eq.~(8.970.1)]{grand00}, we can simplify~\eqref{eq:condPEPNrelays} as
\begin{align}
\label{eq:condPEPNrelays1}
&P^{x_p,x_q}_e(h_{s,d},h_{r_m,d})=e^{\left(\frac{\beta}{4}-\frac{3|\bar{x}|^2}{2}\right)\gamma_t}\sum^\infty_{k=0}\sum^{k+N}_{n=0}\sum^{n}_{i=0}{}^{N+n}C_{n-i}\nonumber\\
&\hspace*{7em}\times\frac{(2|\bar{x}|^2-\beta)^k\left({2|\bar{x}|^2+\beta}\right)^i\gamma^{k+i}_t}{2^{N+n+k+2i+1} k!i!}.
\end{align}
The average uncoded PEP of the proposed PL decoder with $M$-PSK constellation and $N$ error-free relays can be obtained by averaging~\eqref{eq:condPEPNrelays1} over $\gamma_t$. From~\eqref{eq:Xisqpdfgen},~\eqref{eq:condPEPNrelays1}, and~\cite[Eq.~(3.381.4)]{grand00}, the average PEP will be
\begin{align}
\label{eq:avgPEPNrelays}
&P^{x_p,x_q}_e=\frac{1}{\Gamma(N+1)\bar{\gamma}^{N+1}}
\sum^\infty_{k=0}\sum^{k+N}_{n=0}\sum^{n}_{i=0}{}^{N+n}C_{n-i}\nonumber\\
&\hspace*{1em}\times\frac{(2|\bar{x}|^2-\beta)^k\left({2|\bar{x}|^2+\beta}\right)^i}{2^{N+n+k+2i+1} k!i!}\frac{\Gamma(N+k+i+1)}{(\frac{1}{\bar{\gamma}}+c)^{N+k+i+1}},
\end{align}\\
where $c=\frac{3|\bar{x}|^2}{2}-\frac{\beta}{4}$. The average asymptotic approximate SER of the proposed PL decoder with multiple relays can be obtained from~\eqref{eq:serqpsk3} and~\eqref{eq:avgPEPNrelays}.
It can be seen from~\eqref{eq:avgPEPNrelays} that\vspace*{0.5em}
\begin{align}
\label{eq:div}
P^{x_p,x_q}_e\propto \frac{1}{\bar{\gamma}^{N+1}}\frac{1}{(\frac{1}{\bar{\gamma}}+c)^{N+k+i+1}}.
\end{align}\\
Hence, as $\bar{\gamma}\rightarrow \infty$,
\begin{align}
\label{eq:div1}
P^{x_p,x_q}_e\propto \frac{1}{\bar{\gamma}^{N+1}}.
\end{align}\\
Therefore, the proposed PL decoder achieves diversity of $N+1$ in a DF based differential cooperative system with $N$ \emph{error-free} relays. Let $N=1$, then it can be seen from~\eqref{eq:div1} that the PL decoder with a single error-free relay achieves the second order diversity. We have analytically proved in Subsection~\ref{sub:approxhgsnr} that the PL decoder with a single erroneous relay also achieves the second order diversity. Therefore, the proposed PL decoder avoids loss in the diversity because of an erroneous relaying node.  
It is shown in Subsection~\ref{sub:analsym} and Fig.~\ref{fig:asym} by simulations that the PL decoder with $N>1$ \emph{erroneous} relays also achieves diversity of $N+1$. 
\section{Analytical and Simulation Results}
\label{sec:simulation results}
Simulations are performed with $M$-PSK and $M$-QAM constellations. The channels of all links are assumed to be Rayleigh fading and constant over multiple consecutive time intervals. 
\subsection{Performance of the Proposed Decoders for Unitary and Non-Unitary Constellations}
\label{sub:perdf}
\begin{figure}[t!]\vspace*{-1.0em}
  \begin{center}\hspace*{-1.0em}
    \psfig{figure=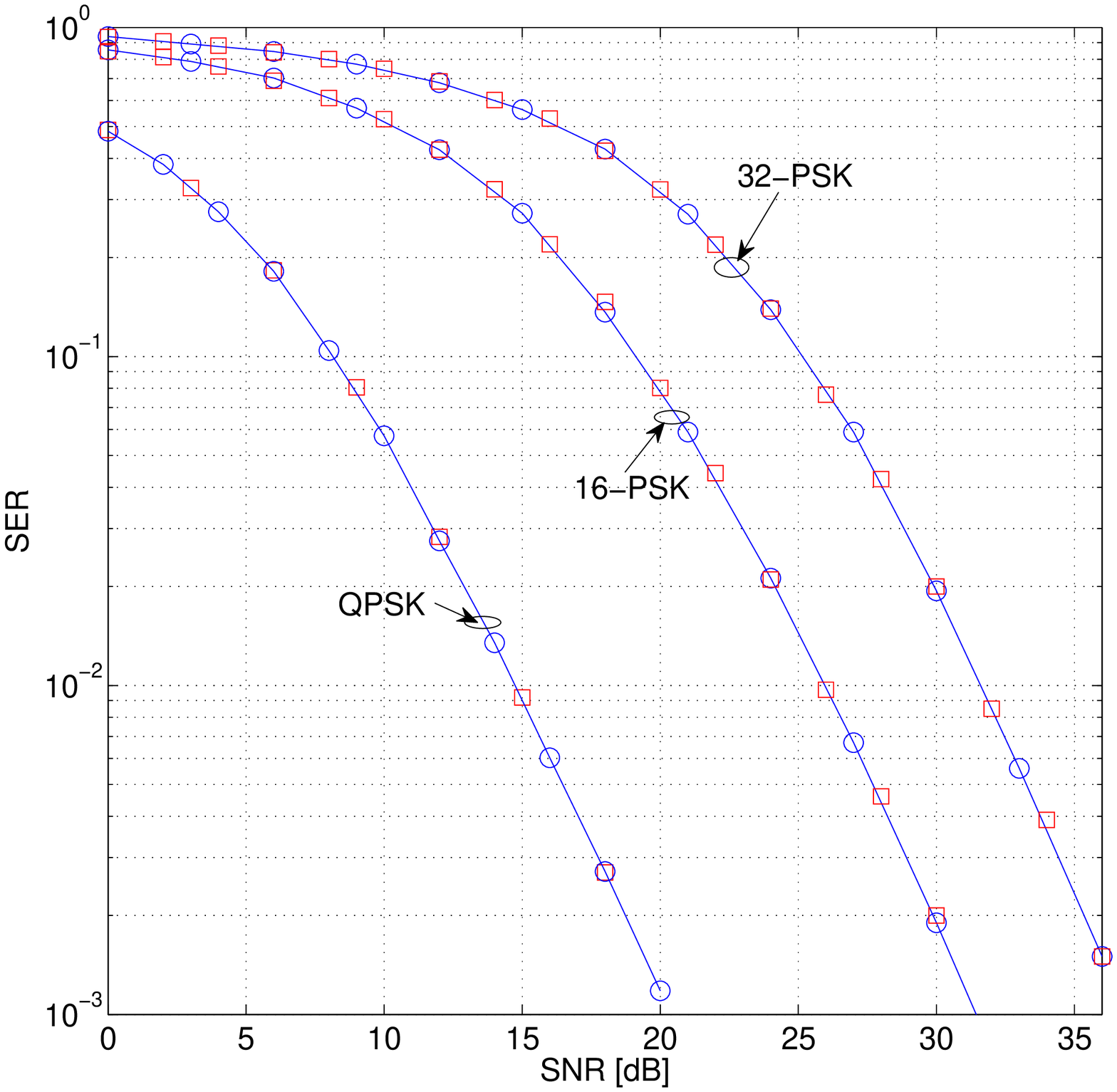,height=2.75in,width=3.75in}
    \vspace*{-0.5em}
    \mycaption{SER versus SNR performance of the proposed ML {\large{${\circ}$}} and PL {{$\square$}} decoders for different $M$-PSK constellations.}
    \label{fig:mlplpsk}
    \vspace*{-0.0em}
  \end{center}
\end{figure}

\begin{figure}[t!]\vspace*{-1.0em}
  \begin{center}\hspace*{-1.0em}
    \psfig{figure=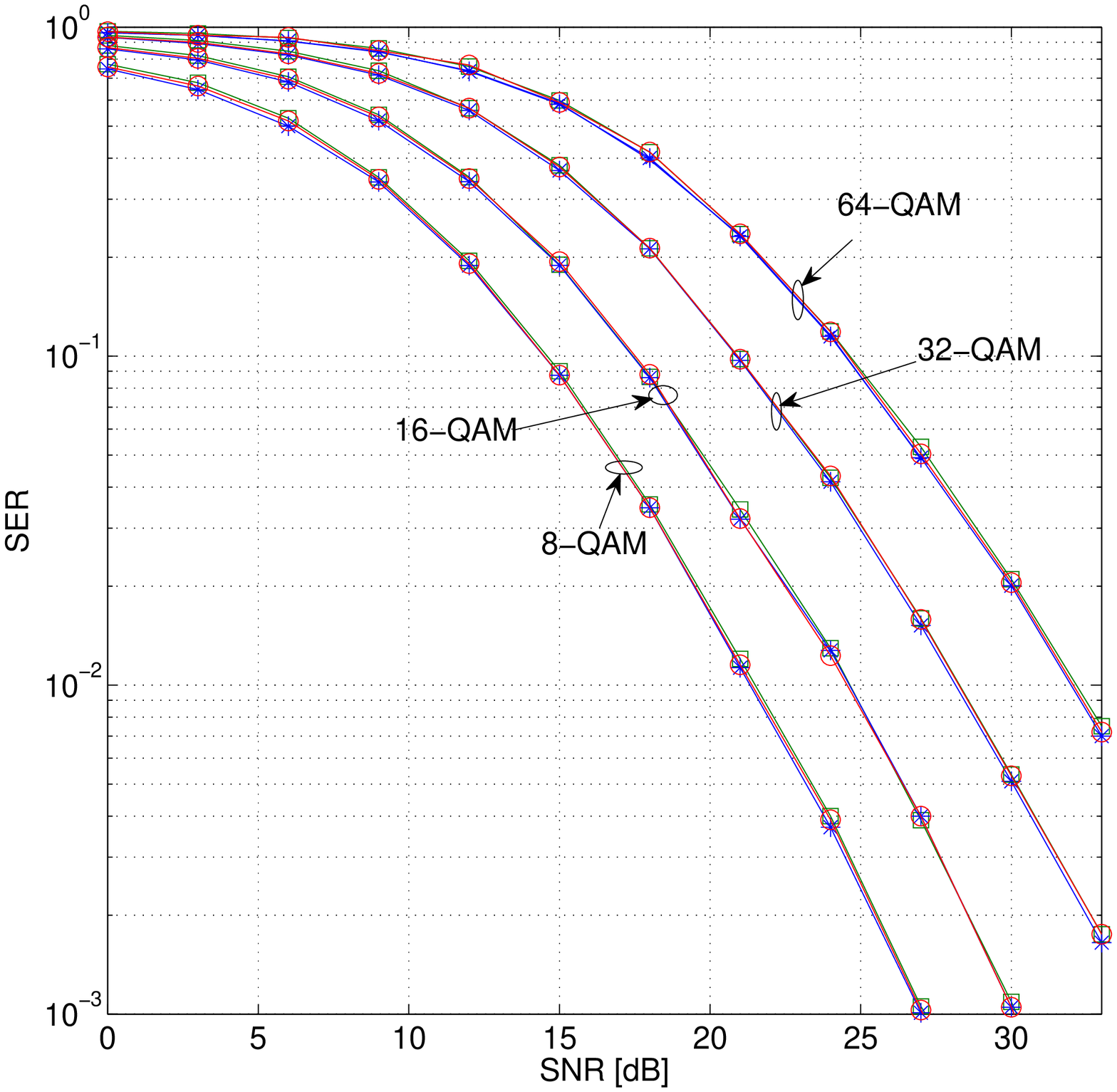,height=2.75in,width=3.75in}
    \vspace*{-0.5em}
\mycaption{SER versus SNR performance of the proposed ML {\large{${\circ}$}} and PL $\square$ decoders utilizing the estimated values of the previously transmitted symbols in the relay and the destination, and the genie added ML decoder {\large{${\ast}$}} for different $M$-QAM constellations.}
    \label{fig:mlplqam}
    \vspace*{-1.0em}
  \end{center}
\end{figure}

\begin{figure}[t!]\vspace*{-1.0em}
  \begin{center}\hspace*{-1.0em}
    \psfig{figure=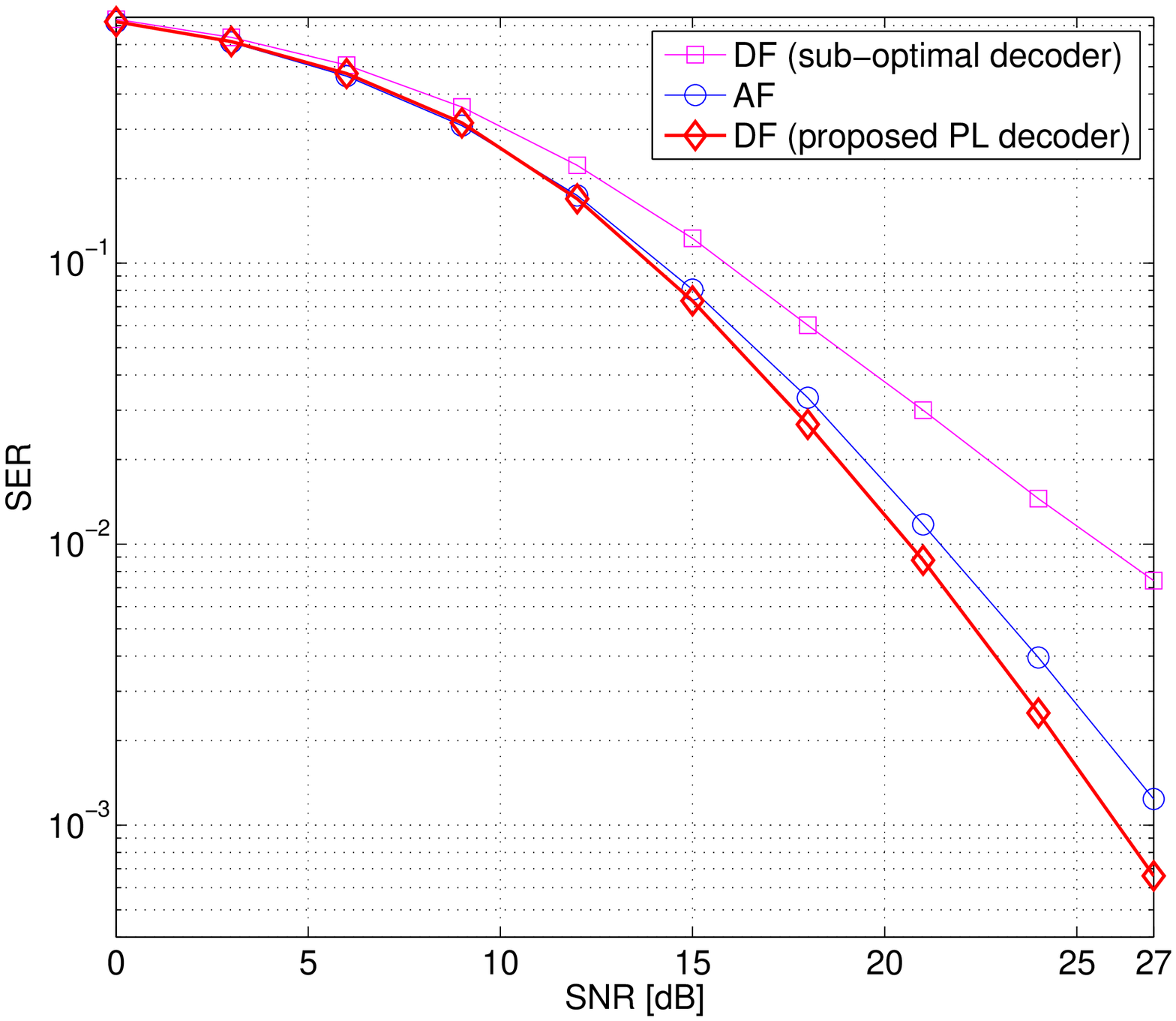,height=2.75in,width=3.75in}
    \vspace*{-0.5em}
\mycaption{Comparison of differential cooperative systems utilizing DF protocol with a sub-optimal decoder and proposed PL decoder, and AF protocol~\cite{himso05} with 8-PSK constellation.}
    \label{fig:afdf}
    \vspace*{0.0em}
  \end{center}
\end{figure}

It is assumed that $\bar{\gamma}_{s,d}=\bar{\gamma}_{s,r}=\bar{\gamma}_{r,d}=\bar{\gamma}$, i.e., all links involved in cooperation have equal average SNR value. We have shown $\bar{\gamma}$ on x-axis in Figs.~\ref{fig:mlplpsk}-~\ref{fig:afdf}. Moreover, we have considered a DF based uncoded cooperative communication system with a \emph{single} relay and one source-destination pair for simulation results shown in Figs.~\ref{fig:mlplpsk}-~\ref{fig:afdf}. 
In Fig.~\ref{fig:mlplpsk}, we have plotted the performance of the proposed ML~\eqref{eq:MLDDest} and PL~\eqref{eq:LLRDDest4} decoders for differential QPSK, 16-PSK, and 32-PSK constellations. It can be seen from Fig.~\ref{fig:mlplpsk} that the ML and PL decoders work approximately similar for all constellations and SNR values considered in the simulations. We have plotted the performance of the proposed ML and PL decoders by using the estimated values of $x[n-1]$ and $x_r[n-1]$ in~\eqref{eq:qam_decoder_relay1},~\eqref{eq:MLDDestqam}, and~\eqref{eq:PL_qam} for differential 8-QAM, 16-QAM, 32-QAM, and 64-QAM constellations in Fig.~\ref{fig:mlplqam}. It can be seen from Fig.~\ref{fig:mlplqam} that the proposed ML and PL decoders also work approximately similar for all $M$-QAM constellations considered in simulations. The SER versus SNR performance of the proposed ML decoder by assuming that $x[n-1]$ is perfectly known in the relay and $x[n-1]$ and $x_r[n-1]$ are perfectly known in the destination is also plotted in Fig.~\ref{fig:mlplqam}. We call this decoder as `genie added ML decoder'. It can be seen from Fig.~\ref{fig:mlplqam} that there is no error propagation in the performance of the proposed ML and PL decoders due to utilization of the estimated values of the previously transmitted symbols.

A cooperative set-up with a single \emph{erroneous} relay and one source-destination pair is considered in Fig.~\ref{fig:afdf} for simulations. In Fig.~\ref{fig:afdf}, we have plotted the SER versus SNR performance of the differentially modulated 8-PSK constellation in the DF based uncoded cooperative system with the proposed PL decoder having knowledge of the instantaneous SNR of the source-relay link and AF based uncoded cooperative system~\cite{himso05}. The total average transmit power per time interval is kept the same in the DF and AF schemes. It can be seen from  Fig.~\ref{fig:afdf} that the uncoded DF based differential cooperative system with the proposed PL decoder outperforms the \emph{same rate} AF based differential cooperative system utilizing uncoded transmissions. For example, a SNR gain of approximately 1~dB is obtained by the differential DF system as compared to the differential AF system at SER=$10^{-2}$.  
Full-diversity ML and PL decoders of higher order unitary and non-unitary constellations in the differential DF system are not available in literature. Therefore, we have plotted the performance of a sub-optimal decoder\footnote{The sub-optimal decoder can be obtained by putting $\epsilon=0$ in~\eqref{eq:MLDDest}.} that does not have any information of the SNR of the source-relay link in Fig.~\ref{fig:afdf}. Since the sub-optimal decoder does not have knowledge of the SNR of the source-relay link, it \emph{wrongly} assumes that the relay is error-free, whereas, the relay actually performs erroneous transmissions. It can be seen from Fig.~\ref{fig:afdf} that the sub-optimal decoder performs poorer to the proposed PL decoder at all SNRs considered in the figure. Moreover, the sub-optimal decoder looses diversity due to the erroneous transmissions of the relaying node. 
\subsection{Analytical Performance of the Proposed PL Decoder}
\label{sub:analsym}
\begin{figure}[t!]\vspace*{-1.0em}
  \begin{center}\hspace*{-1.0em}
    \psfig{figure=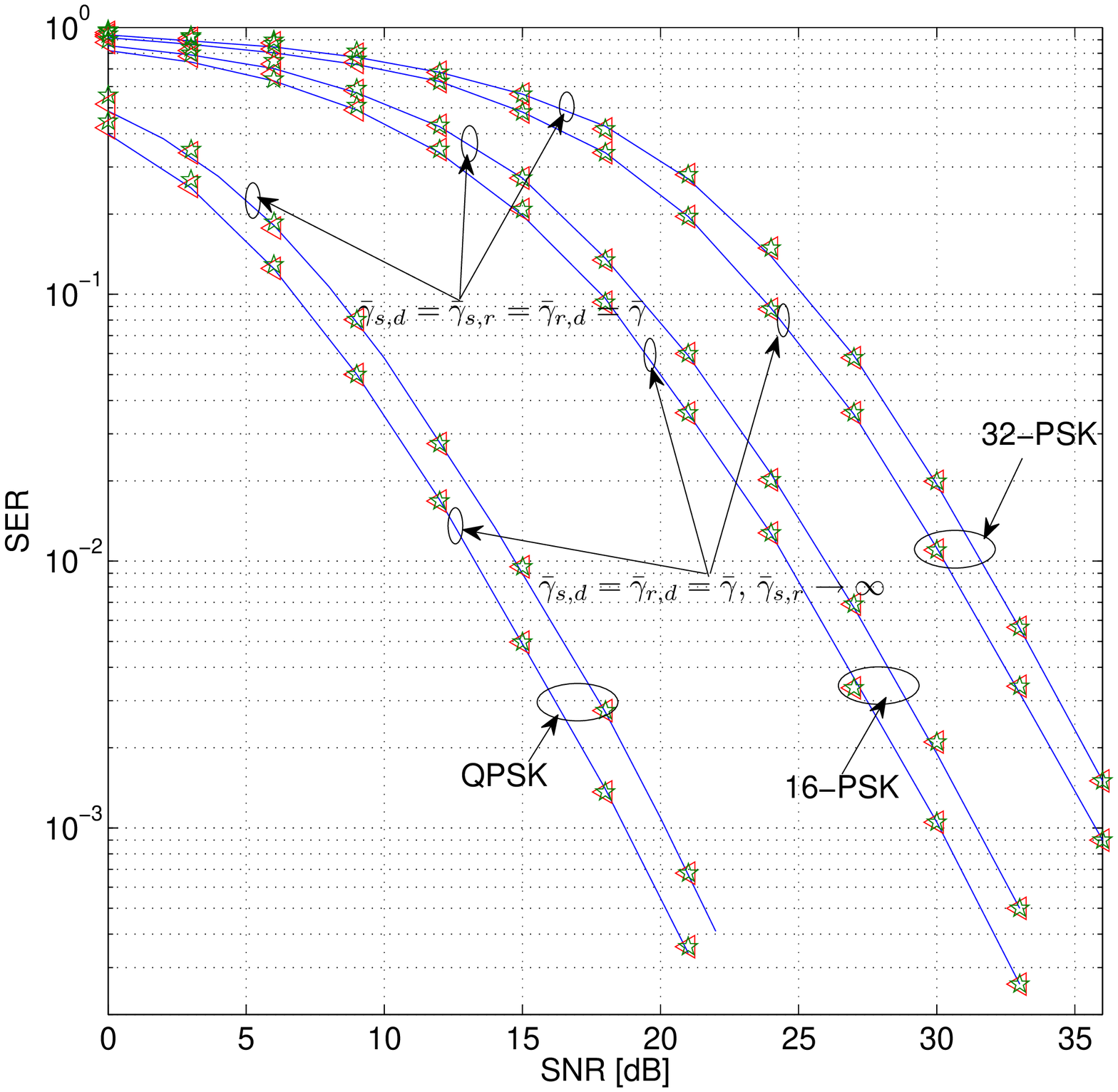,height=2.75in,width=3.75in}
    \vspace*{-0.5em}
\mycaption{Analytical ($\triangle$ SER obtained in closed-form by using~\eqref{eq:avgser1} and~\eqref{eq:serqpsk3} and {\large ${\star}$} SER obtained by numerically solving the integrals from~\eqref{eq:serqpsk3} and~\eqref{eq:pep}, and ignoring the higher order noise) and simulated {\large$-\!\!-$} performance of the proposed PL decoder with $\bar{\gamma}_{s,d}=\bar{\gamma}_{r,d}=\bar{\gamma}_{s,r}=\bar{\gamma}$ and $\bar{\gamma}_{s,d}=\bar{\gamma}_{r,d}=\bar{\gamma},\bar{\gamma}_{s,r}\rightarrow \infty$, where 0~dB$\leq\bar{\gamma}\leq 36$~dB. An uncoded cooperative system with a single relay is considered in the simulations and analysis.}
    \label{fig:analysis}
    \vspace*{0.0em}
  \end{center}
\end{figure}

We have plotted the analytical approximate average SER versus SNR plots of the QPSK, 16-PSK, and 32-PSK constellations in the differential cooperative communication system having one source-destination pair and a single relay in Fig.~\ref{fig:analysis}. It is assumed that $\bar{\gamma}_{s,d}=\bar{\gamma}_{s,r}=\bar{\gamma}_{r,d}=\bar{\gamma}$. The approximate values of the average SER are calculated in closed-form by using~\eqref{eq:avgser1} and~\eqref{eq:serqpsk3}, and by numerically solving the integrals in~\eqref{eq:pep} and then using~\eqref{eq:serqpsk3}. It can be seen from Fig.~\ref{fig:analysis} that the simulation results follow the analytical results satisfactorily at all SNR values. Moreover, there is no significant degradation in the analysis by ignoring the higher order noise terms for all constellations considered in the figure. 
We have also plotted the simulated and analytical performance of the differential cooperative system with one relay under the condition that $\bar{\gamma}_{s,d}=\bar{\gamma}_{r,d}=\bar{\gamma}$, $\bar{\gamma}_{s,r}\rightarrow \infty$, i.e., the channel between the source and relay is error-free for QPSK, 16-PSK, and 32-PSK constellations, in Fig.~\ref{fig:analysis}. 
It can be seen from Fig.~\ref{fig:analysis} that the proposed analysis closely justifies the simulated behavior of the DF based differential cooperative system with error-free relay. 

\begin{figure}[t!]\vspace*{-1.0em}
  \begin{center}\hspace*{-1.0em}
    \psfig{figure=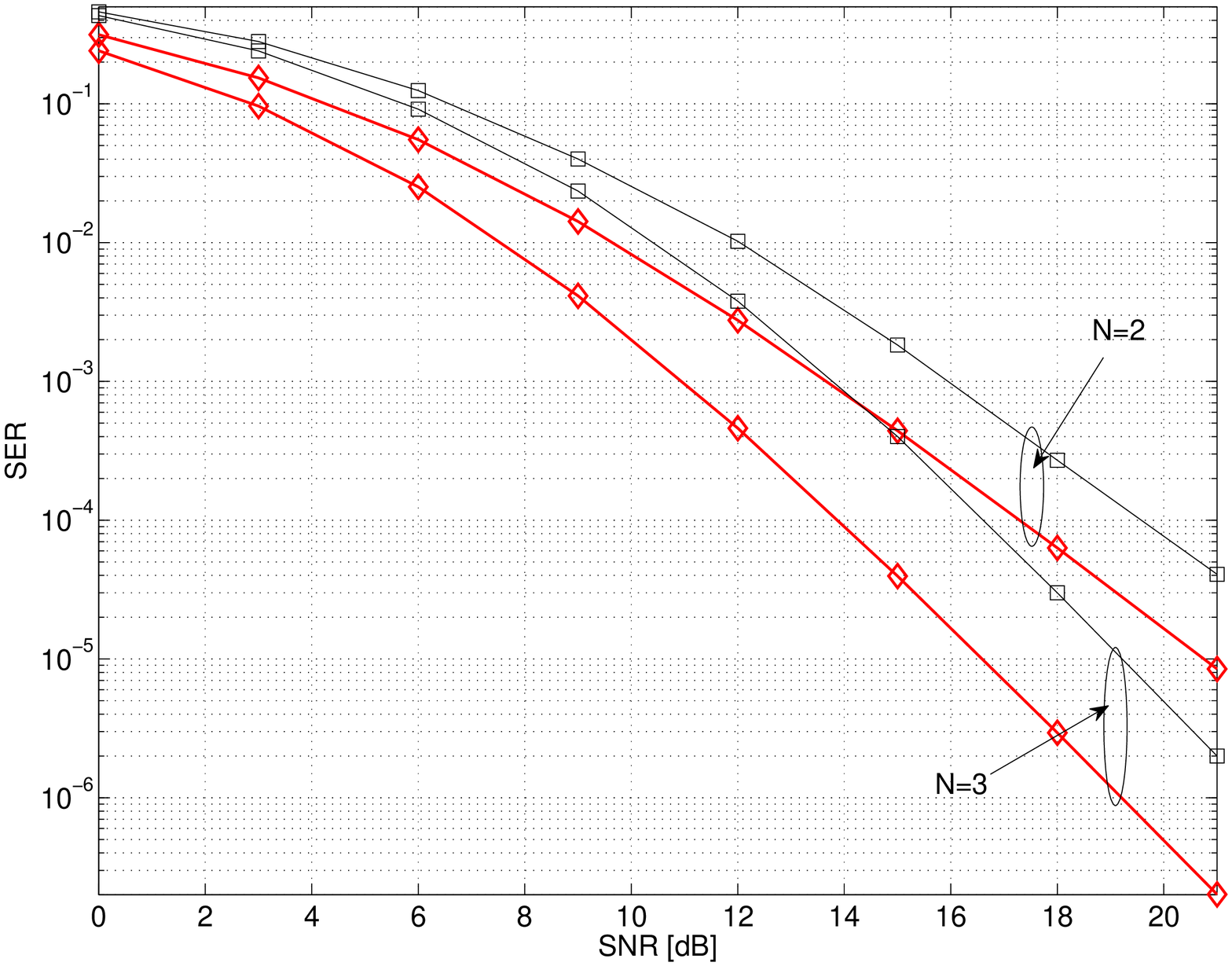,height=2.75in,width=3.75in}
    \vspace*{-0.5em}
\mycaption{Analytical {\large{$-\!\!\diamond\!\!-$}} performance of the proposed PL decoder with  $\bar{\gamma}_{s,d}=\bar{\gamma}_{r,d}=\bar{\gamma},\bar{\gamma}_{s,r}\rightarrow \infty$ and simulated $-\!\square\!-$ performance of the proposed PL decoder with  $\bar{\gamma}_{s,d}=\bar{\gamma}_{r,d}=\bar{\gamma}_{s,r}=\bar{\gamma}$, where 0~dB$\leq\bar{\gamma}\leq 21$~dB and $N=2,3$.}
    \label{fig:asym}
    \vspace*{-0.5em}
  \end{center}
\end{figure}

In Fig.~\ref{fig:asym}, we have plotted the analytical approximate asymptotic SER of the differential cooperative system with a single source-destination pair, $N=2,3$ relays, QPSK constellation, $\bar{\gamma}_{s,d}=\bar{\gamma}_{r,d}=\bar{\gamma}$, and $\bar{\gamma}_{s,r}\rightarrow \infty$. 
The analytical asymptotic SER versus SNR values are calculated from~\eqref{eq:serqpsk3} and~\eqref{eq:avgPEPNrelays}. We have also plotted the simulated SER versus SNR plots of the proposed PL decoder for $N=2,3$, QPSK constellation, and $\bar{\gamma}_{s,d}=\bar{\gamma}_{s,r}=\bar{\gamma}_{r,d}=\bar{\gamma}$. It can be noticed that the analytical asymptotic SER obtained from~\eqref{eq:serqpsk3} and~\eqref{eq:avgPEPNrelays} is a lower bound of the SER of the PL decoder with erroneous relays. 
From Fig.~\ref{fig:asym}, it can be seen that for a fixed number of relays, the SER versus SNR plot of the PL decoder with erroneous relays decays at the same rate as that of the PL decoder with error-free relays at high SNR considered in the figure. The diversity is defined as slope of the SER versus SNR plot~\cite{larss03}. Therefore, from Fig.~\ref{fig:asym}, it can be noticed that the proposed PL decoder for a given number ($N=2,3$) of error-free and erroneous relays achieves the same diversity. We have analytically proved in Subsection~\ref{sub:asyanal} that the diversity (slope of decay of the SER versus SNR plot) of the proposed PL decoder with $N$ error-free relays is $N+1$. Therefore, from Fig.~\ref{fig:asym} and the discussion above, it can be noticed that the proposed PL decoder with $N>1$ erroneous relays also achieves diversity of $N+1$.
The x-axis in Fig.~\ref{fig:asym} depicts the average SNR of the source-destination link. . 
\section{Conclusions}
We have derived optimal and low complexity decoders for differentially modulated complex-valued constellations in an uncoded cooperative communication system utilizing the DF protocol and multiple relays. Moreover, we have also derived expressions of  the approximate average symbol error rate of the proposed PL decoder. 
It is proved by analysis and simulations that the proposed PL decoder achieves the maximum possible diversity.   
\appendices
\section{Derivation of Uncoded Average PEP with A Single Relay}
\label{app:condPEP}
By considering the erroneous decoding in the relay, we can write
\begin{align}
\label{eq:term1}
\text{Pr}\left\{t_0-T\right.&\left.<0|t<-T,x[n]=x_p\right\}\text{Pr}\left\{t<-T|x[n]=x_p\right\}\nonumber\\
&=\text{Pr}\left\{t_0-T<0|t<-T,x[n]=x_p\right\}\nonumber\\
&\times\left((1-\epsilon)\text{Pr}\left\{t<-T|x[n]=x_p,x_r[n]=x_p\right\}\right.\nonumber\\
&\hspace*{0em}\left.+\epsilon\text{Pr}\left\{t<-T|x[n]=x_p,x_r[n]\neq x_p\right\}\right)\nonumber\\
&=P_{e_1}(h_{s,d})\left(P_{e_2}(h_{r,d})+P_{e_3}(h_{r,d})\right).
\end{align}
Similarly, we have
\begin{align}
\label{eq:term2}
\text{Pr}\left\{t_0+T\right.&\left.<0|t>T,x[n]=x_p\right\}\text{Pr}\left\{t>T|x[n]=x_p\right\}\nonumber\\
&=\text{Pr}\left\{t_0+T<0|t>T,x[n]=x_p\right\}\nonumber\\
&\times\left((1-\epsilon)\text{Pr}\left\{t>T|x[n]=x_p,x_r[n]=x_p\right\}\right.\nonumber\\
&\hspace*{0em}\left.+\epsilon\text{Pr}\left\{t>T|x[n]=x_p,x_r[n]\neq x_p\right\}\right)\nonumber\\
&=P_{e_4}(h_{s,d})\left(P_{e_5}(h_{r,d})+P_{e_6}(h_{r,d})\right).
\end{align}\\
For given $h_{s,d}$, $h_{r,d}$, $x[n]$, and $x_r[n]$, $t_0$ and $t$ follow quadratic form of Gaussian variates. Therefore, 
from~\cite[Section~IV]{biyar93} we have $P_{e_1}(h_{s,d})=1-g(c_p,b_p,\gamma_{s,d})$, $P_{e_2}(h_{r,d})=(1-\epsilon)g(b_p,c_p,\gamma_{r,d})$, $P_{e_4}(h_{s,d})=g(b_p,c_p,\gamma_{s,d})$, and $P_{e_5}(h_{r,d})=(1-\epsilon)g(c_p,b_p,\gamma_{r,d})$, where
\begin{align}
\label{eq:intfunc}
g(a_i,a_j,\gamma_b)\triangleq&\frac{1}{2}e^{-\gamma_b(2|\bar{x}|^2-\frac{a_i}{8})}\sum^{\infty}_{k=0}\sum^k_{n=0}\frac{\gamma^k_b a^k_j}{4^k k!} \nonumber\\
&\times \frac{\Gamma(k-n+1,2T)}{2^n (k-n)!}L_n(-\frac{a_i\gamma_b}{8}),
\end{align}\\
$a_i, a_j,$ and $\gamma_b$ are variables, $\Gamma(\cdot,\cdot)$ denotes the incomplete Gamma function~\cite[Eq.~(6.5.3)]{abram72}, $L_n(\cdot)$ is the Laguerre polynomial~\cite[pg.~775]{abram72}, $b_p=2(2|\bar{x}|^2+x^*_p\bar{x}+x_p\bar{x}^*)$, and $c_p=2(2|\bar{x}|^2-x^*_p\bar{x}-x_p\bar{x}^*)$. 

Whereas, $\text{Pr}\left\{t<-T|x[n]=x_p,x_r[n]\neq x_p\right\}$ leads to the cumulative distribution function (c.d.f.) of the quadratic Gaussian mixture random variable which can be obtained by marginalizing the c.d.f. of the quadratic Gaussian random variable~\cite[Eq.~(27)]{biyar93} over $x_i, i\neq p$ as follows:
\begin{align}
\label{eq:quadmixgauss}
\text{Pr}&\left\{t<-T|x[n]=x_p,x_r[n]\neq x_p\right\}\nonumber\\
&=\frac{1}{2(M-1)}\sum^{M}_{i=0\atop i\neq p}\sum^{\infty}_{k=0}\sum^k_{n=0}e^{-\gamma_{r,d}(2|\bar{x}|^2-\frac{b_i}{8})}\frac{\gamma^k_{r,d} c^k_i}{4^k k!}\nonumber\\
&\hspace*{3em}\times \frac{\Gamma(k-n+1,2T)}{2^n (k-n)!}L_n(-\frac{b_i\gamma_{r,d}}{8}),
\end{align}\\
where $b_i=2(2|\bar{x}|^2+x^*_i\bar{x}+x_i\bar{x}^*)$ and $c_i=2(2|\bar{x}|^2-x^*_i\bar{x}-x_i\bar{x}^*)$.
From~\eqref{eq:term1},~\eqref{eq:intfunc} and~\eqref{eq:quadmixgauss}, we have $P_{e_3}(h_{r,d})=\frac{\epsilon}{M-1}\displaystyle \sum\limits^M_{i=1\atop i\neq p} g(b_i,c_i,\gamma_{r,d})$. 
Similarly, we can obtain $P_{e_6}(h_{r,d})=\frac{\epsilon}{M-1}\displaystyle \sum\limits^M_{i=1\atop i\neq p} g(c_i,b_i,\gamma_{r,d})$. Next, we can write
\begin{align}
\label{eq:term3}
&\text{Pr}\left\{t_0+t<0,-T\leq t\leq T|x[n]=x_p\right\}\nonumber\\
&=(1-\epsilon)\text{Pr}\left\{t_0+t<0,-T\leq\!\! t\!\!\leq T|x[n]\!\!=\!\!x_p,x_r[n]\!\!=\!\!x_p\right\}\nonumber\\
&+\epsilon\text{Pr}\left\{t_0+t<0,-T\leq t\leq T|x[n]=x_p,x_r[n]\neq x_p\right\}.
\end{align}
It can be shown after some algebra that
\begin{align}
\label{eq:jointterm1}
\text{Pr}&\left\{t_0+t<0,-T\leq t\leq T|x[n]=x_p,x_r[n]=x_p\right\}\nonumber\\
&=\int^{T}_{-T}p_{t|x_r[n]=x_p}(w)\int^{-w}_{-\infty}p_{t_0}(z) dz\:dw\nonumber\\
&=\int^{0}_{-T}p_{t|x_r[n]=x_p}(w)\int^{-w}_{-\infty}p_{t_0}(z) dz\:dw\nonumber\\
&+\int^{T}_{0}p_{t|x_r[n]=x_p}(w)\int^{-w}_{-\infty}p_{t_0}(z) dz\:dw\nonumber\\
&=\int^{0}_{-T}p_{t|x_r[n]=x_p}(w)F_{t_0}(-w) dw\nonumber\\
&+\int^{T}_{0}p_{t|x_r[n]=x_p}(w)F_{t_0}(-w) dw\nonumber\\
&=\frac{1}{1-\epsilon}P_{e_7}(h_{s,d},h_{r,d})+\frac{1}{1-\epsilon}P_{e_8}(h_{s,d},h_{r,d}),
\end{align}\\
where $F_{t_0}(x)$ is the c.d.f. of $t_0$ and $p_{t|x_r[n]=x_p}(x)$ is the p.d.f. of $t$ given that $x_r[n]=x_p$. By change of variable in~\eqref{eq:jointterm1}, using~\cite[Eqs.~(15),(27),(29)]{biyar93}, series expansion of the incomplete Gamma function
\begin{align}
\label{eq:gammaseries}
\Gamma(v,y)=(v-1)!e^{-y}\sum^{v-1}_{k=0}\frac{y^k}{k!},
\end{align}\\
and then using~\cite[Eq.~(3.381.1)]{grand00}, we can obtain $P_{e_7}(h_{s,d},h_{r,d})=(1-\epsilon)\displaystyle\sum\limits_{k,n,m_1,\atop l,i_1}^{}\mathcal{D}^{c_p,b_p}_{k,n}\mathcal{D}^{b_p,c_p}_{m_1,l}\mathcal{D}^{k,n}_{i_1,m_1,l}$ and $P_{e_8}(h_{s,d},h_{r,d})=\!(1-\epsilon)\left[\!\displaystyle\sum\limits^\infty_{k=0}\!\displaystyle\sum\limits^k_{n=0}\!\mathcal{D}^{b_p,c_p}_{k,n}2^{n-k-1}\right.\\\left.\times\gamma(-n+k+1,2T)-\displaystyle\sum\limits_{k,n,m_1,\atop l,i_1}^{}\mathcal{D}^{b_p,c_p}_{k,n}\mathcal{D}^{c_p,b_p}_{m_1,l}\mathcal{D}^{k,n}_{i_1,m_1,l}\right]$, where $\mathcal{D}^{a_i,a_j}_{k,n}=e^{-\gamma_{r,d}\left(2|\bar{x}|^2-\frac{a_i}{8}\right)}\frac{a^k_j\gamma^k_{r,d}}{k!2^k}\frac{L_n(-\frac{a_i\gamma_{r,d}}{8})}{(k-n)!4^n}$, $\mathcal{D}^{k,n}_{i_1,m_1,l}=(m_1-l)!\frac{2^{i_1}}{i_1!}4^{n-k-i_1-1}\gamma(k-n+i_1+1,4T)$,
$\mathcal{D}^{a_i,a_j}_{m_1,l}=\frac{1}{2}e^{-\gamma_{s,d}\left(2|\bar{x}|^2-\frac{a_i}{8}\right)}\frac{a^{m_1}_j\gamma^{m_1}_{s,d}}{m_1!4^{m_1}} \frac{L_l(-\frac{a_i\gamma_{s,d}}{8})}{(m_1-l)!2^l}$, 
and $\displaystyle\sum\limits_{k,n,m_1,\atop l,i_1}^{}=\sum^\infty_{k=0}\sum^k_{n=0}\sum^\infty_{m_1=0}\sum^{m_1}_{l=0}\sum^{m_1-l}_{i_1=0}$.
Similarly, we have
\begin{align}
\label{eq:jointterm2}
\text{Pr}&\left\{t_0+t<0,-T\leq t\leq T|x[n]=x_p,x_r[n]\neq x_p\right\}\nonumber\\
&=\int^{0}_{-T}p_{t|x_r[n]\neq x_p}(w)F_{t_0}(-w) dw\nonumber\\
&+\int^{T}_{0}p_{t|x_r[n]\neq x_p}(w)F_{t_0}(-w) dw\nonumber\\
&=\frac{1}{\epsilon}P_{e_9}(h_{s,d},h_{r,d})+\frac{1}{\epsilon}P_{e_{10}}(h_{s,d},h_{r,d}),
\end{align}\\
where $p_{t|x_r[n]\neq x_p}(x)$ is the p.d.f. of $t$ given that $x_r[n]\neq x_p$. Since $t$ is the quadratic Gaussian mixture random variable, the p.d.f. of $t$ can be obtained by marginalizing $p_{t|x_r[n]=x_i}(x)$ over $x_i, i\neq p$ as
\begin{align}
\label{eq:pdfgaussmix}
p_{t|x_r[n]\neq x_p}(v)=\left\{\begin{array}{c}\frac{1}{M-1}\displaystyle\sum\limits^{M}_{i=1 \atop i\neq p}\sum^{\infty}_{k=0}\sum^{k}_{n=0} e^{-\left(2v+\frac{|\bar{x}|^2\gamma_{r,d}}{2}+\frac{c_i\gamma_{r,d}}{8}\right)}\\
\times\frac{v^{k-n}\gamma^k_{r,d}b^k_i}{k!(k-n)!2^{n+k}} L_n\left(-\frac{c_i\gamma_{r,d}}{8}\right),\ms v>0,\\
\frac{1}{M-1}\displaystyle\sum\limits^{M}_{i=1 \atop i\neq p}\sum^{\infty}_{k=0}\sum^{k}_{n=0}e^{-\left(2v+\frac{|\bar{x}|^2\gamma_{r,d}}{2}+\frac{b_i\gamma_{r,d}}{8}\right)}\\ \times\frac{(-v)^{k-n}\gamma^k_{r,d}c^k_i}{k!(k-n)!2^{n+k}} L_n\left(-\frac{b_i\gamma_{r,d}}{8}\right),\ms v\leq 0.
\end{array}\right.
\end{align} \\
From~\eqref{eq:gammaseries},~\eqref{eq:jointterm2},~\eqref{eq:pdfgaussmix},~\cite[Eqs.~(27),(29)]{biyar93}, and~\cite[Eq.~(3.381.1)]{grand00} we can obtain 
$P_{e_9}(h_{s,d},h_{r,d})=\frac{\epsilon}{M-1}\displaystyle\sum\limits^M_{i=1\atop i\neq p}\displaystyle\sum\limits_{k,n,m_1,\atop l,i_1}^{}\mathcal{D}^{c_i,b_i}_{k,n}\mathcal{D}^{b_p,c_p}_{m_1,l}\mathcal{D}^{k,n}_{i_1,m_1,l}$ and
$P_{e_{10}}(h_{s,d},h_{r,d})=\frac{\epsilon}{M-1}\left[\displaystyle\sum\limits^M_{i=1\atop i\neq p}\sum^\infty_{k=0}\sum^k_{n=0}\mathcal{D}^{b_i,c_i}_{k,n}2^{n-k-1}\right.\gamma(-n+k+1,2T_m)\\\left.
-\displaystyle\sum\limits^M_{i=1\atop i\neq p}\displaystyle\sum\limits_{k,n,m_1,\atop l,i_1}^{}\mathcal{D}^{b_i,c_i}_{k,n}\mathcal{D}^{c_p,b_p}_{m_1,l}\mathcal{D}^{k,n}_{i_1,m_1,l}\right]$, where
$\gamma\left(\cdot,\cdot\right)$ is the incomplete Gamma function~\cite[Eq.~(6.5.2)]{abram72}.
 
A close examination of the terms $P_{e_i}\left(\cdot\right)$ and $P_{e_j}\left(\cdot,\cdot\right)$ of~\eqref{eq:condPEP} reveals that the 2-D integration in~\eqref{eq:avgser} is separable. After some algebra and using~\cite[Eqs.~(3.381.4) and~(8.970.1)]{grand00}, we have 
$P_{e_1}=1-g_1(b_p,c_p,\bar{\gamma}_{s,d})$, $P_{e_2}=(1-\epsilon)g_1(c_p,b_p,\bar{\gamma}_{s,d})$, $P_{e_3}\\=\frac{\epsilon}{M-1}\displaystyle\sum\limits^{M}_{i=1\atop i\neq p}g_1(c_i,b_i,\bar{\gamma}_{s,d})$, $P_{e_4}=g_1(c_p,b_p,\bar{\gamma}_{s,d})$, $P_{e_5}=(1-\epsilon)g_1(b_p,c_p,\bar{\gamma}_{s,d})$, $P_{e_6}=\frac{\epsilon}{M-1}\displaystyle\sum\limits^{M}_{i=1\atop i\neq p}g_1(b_i,c_i,\bar{\gamma}_{s,d})$,
$g_1(a,b,\bar{\gamma})\!\!=\!\!\frac{1}{2\bar{\gamma}}\!\!\displaystyle\sum\limits^\infty_{k=0}\!\sum^k_{n=0}\!\sum^n_{i_2=0}\frac{a^k}{k!4^k}\frac{\Gamma(k-n+1,2T_m)}{(k-n)!2^n}{}^nC_{i_2}\frac{b^{i_2}}{i_2!8^{i_2}}(2|\bar{x}|^2-\frac{b}{8}\\+\frac{1}{\bar{\gamma}})^{-i_2-k-1}\Gamma(i_2+k+1)$,
$P_{e_7}=(1-\epsilon)\displaystyle\sum\limits_{k,n,m_1,l,\atop i_1,i_2,i_3}^{}\mathcal{B}^{c_p,b_p}_{k,n,i_2}\\\times\mathcal{B}^{b_p,c_p}_{m_1,l,i_3}\mathcal{D}^{k,n}_{i_1,m_1,l}$,
$
P_{e_8}=(1-\epsilon)\left[\displaystyle\sum\limits^\infty_{k=0}\sum^k_{n=0}\sum^{n}_{i_2=0}\mathcal{B}^{b_p,c_p}_{k,n,i_2}2^{n-k-1}\right.\\\left.\times\gamma(k-n+1,2T_m)-\displaystyle\sum\limits_{k,n,m_1,l,\atop i_1,i_2,i_3}^{}\mathcal{B}^{b_p,c_p}_{k,n,i_2}\mathcal{B}^{c_p,b_p}_{m_1,l,i_3}\mathcal{D}^{k,n}_{i_1,m_1,l}\right]$,
$P_{e_9}\!\!\!=\!\!\frac{\epsilon}{M-1}\!\displaystyle\sum\limits^M_{i=1\atop i\neq p}\displaystyle\sum\limits_{k,n,m_1,l,\atop i_1,i_2,i_3}^{}\mathcal{B}^{c_i,b_i}_{k,n,i_2}\mathcal{B}^{b_p,c_p}_{m_1,l,i_3}\mathcal{D}^{k,n}_{i_1,m_1,l}$,
$P_{e_{10}}=\frac{\epsilon}{M-1}\left[\displaystyle\sum\limits^M_{i=1\atop i\neq p}\sum^\infty_{k=0}\sum^k_{n=0}\displaystyle\sum\limits^{n}_{i_2=0}\mathcal{B}^{b_i,c_i}_{k,n,i_2}2^{n-k-1}\gamma(k-n+1,2T)\right.\\\left.\!\!-\!\!\sum^M_{i=1\atop i\neq p}\!\!\displaystyle\sum\limits_{k,n,m_1,l,\atop i_1,i_2,i_3}^{}\mathcal{B}^{b_i,c_i}_{k,n,i_2}\mathcal{B}^{c_p,b_p}_{m_1,l,i_3}\mathcal{D}^{k,n}_{i_1,m_1,l}\right]$, \:\:where
$\displaystyle\sum\limits_{k,n,m_1,l,\atop i_1,i_2,i_3}^{}\!\!\!\!\\=\!\!\!\sum^\infty_{k=0}\!\sum^k_{n=0}\!\sum^\infty_{m_1=0}\!\sum^{m_1}_{l=0}\!\sum^{m_1-l}_{i_1=0}\sum^{n}_{i_2=0}\sum^{l}_{i_3=0}$, 
$\mathcal{B}^{c,b}_{k,n,i_2}=\frac{{}^nC_{i_2}b^k c^{i_2}}{\bar{\gamma}_{r,d}k!2^k(k-n)!i_2!4^n8^{i_2}}(2|\bar{x}|^2-\frac{c}{8}+\frac{1}{\bar{\gamma}_{r,d}})^{-i_2-k-1}\Gamma(i_2+k+1)$, and 
$\mathcal{B}^{b,c}_{m_1,l,i_3}=\frac{1}{2\bar{\gamma}_{s,d}}\frac{{}^lC_{i_3}c^{m_1} b^{i_3}}{m_1!4^{m_1}(m_1-l)!i_3!2^l8^{i_3}}(2|\bar{x}|^2-\frac{b}{8}+\frac{1}{\bar{\gamma}_{s,d}})^{-i_3-m_1-1}\Gamma(i_3+m_1+1)$.
\bibliography{IEEEabrv,biblitt}
\bibliographystyle{IEEEtran}
%
%
\end{document}